\documentclass[journal,twoside,web]{ieeecolor}
\usepackage{tmi}
\usepackage{cite}
\usepackage{amsmath,amssymb,amsfonts}
\usepackage{graphicx}
\usepackage{textcomp}
\def\BibTeX{{\rm B\kern-.05em{\sc i\kern-.025em b}\kern-.08em
    T\kern-.1667em\lower.7ex\hbox{E}\kern-.125emX}}

\usepackage[mathscr]{eucal} 
\usepackage{amsmath,amscd,amssymb,amsbsy}
\usepackage{graphicx}
\usepackage[caption=false,font=footnotesize,farskip=0pt]{subfig}
\usepackage{algorithm}
\usepackage{algorithmic}
\usepackage{verbatim}
\usepackage{array}
\usepackage{cite}
\usepackage{url}
\usepackage{booktabs}
\usepackage{multirow}
\usepackage{float}

\begin{document}
%
\title{Hard Sample Aware Noise Robust Learning for Histopathology Image Classification}

\author{Chuang Zhu,
        Wenkai Chen,
        Ting Peng,
        Ying Wang,
        Mulan Jin
 \thanks{The first two authors contributed equally to this work.}
 \thanks{This work was supported in part by the National Natural Science Foundation of China under Grant 81972248, and in part by the Natural Science Foundation of Beijing Municipality under Grant 7202056.}
 \thanks{C. Zhu, W.K. Chen, and T. Peng are with the School of Artificial Intelligence, Beijing University of Posts and Telecommunications, Beijing 100876, China (Corresponding author: Chuang Zhu (czhu@bupt.edu.cn)).}
 \thanks{Y. W. and M.L. Jin are with Beijing Chaoyang Hospital, Capital Medical University
Beijing, China.}
}

%

\maketitle

\begin{abstract}
Deep learning-based histopathology image classification is a key technique to help physicians in improving the accuracy and promptness of cancer diagnosis. However, the noisy labels are often inevitable in the complex manual annotation process, and thus mislead the training of the classification model. In this work, we introduce a novel hard sample aware noise robust learning method for histopathology image classification. To distinguish the informative hard samples from the harmful noisy ones, we build an easy/hard/noisy (EHN) detection model by using the sample training history. Then we integrate the EHN into a self-training architecture to lower the noise rate through gradually label correction. With the obtained almost clean dataset, we further propose a noise suppressing and hard enhancing (NSHE) scheme to train the noise robust model. Compared with the previous works, our method can save more clean samples and can be directly applied to the real-world noisy dataset scenario without using a clean subset. Experimental results demonstrate that the proposed scheme outperforms the current state-of-the-art methods in both the synthetic and real-world noisy datasets. The source code and data are available at \emph{https://github.com/bupt-ai-cz/HSA-NRL/}.
\end{abstract}


\IEEEpeerreviewmaketitle

\section{Introduction}
\IEEEPARstart{C}{ancer} is a serious threat to people's life and health. The studies \cite{schiffman2015early} confirmed that the early screening of cancer is crucial for enhancing the survival rate. The pathological examination is the golden standard of early cancer detection, which can determine the tissue source, nature, and scope of the tumor relying on the visual observation of pathologists. However, there are still many challenges to overcome. During the actual diagnosis process, pathologists analyze the overall tissue along with nuclei organization, density, and variability, which requires tedious workloads. The diagnosis accuracy can be negatively affected by many factors, such as pathologist fatigue and distraction, and the complexity of the tissue structure \cite{wang2019breast}.

The deep learning (DL) techniques, such as convolutional neural networks (CNNs) have been widely used in the fields of histopathology image analysis\cite{campanella2019clinical-grade,chen2019an}. These efforts are designed to help physicians in improving the accuracy and promptness of cancer diagnosis. One typical task in the field of histopathology image analysis is image classification. Several works \cite{coudray2018classification,tang2019interpretable} are developed to build a deep learning model to classify the histopathology images. However, 
all these works assume that the utilized dataset was clean for model training.

Actually, it is very expensive and difficult to collect a large dataset with clean labels \cite{yi2019probabilistic}. In the real medical diagnosis scenarios, noisy labels are often inevitable in manual annotation due to the following reasons: 1) expert domain knowledge is required to perform labeling; 2) manual annotation suffers from large intra- and inter-observer variability even among experts; 3) it is time-consuming and tedious to annotate a large number of patches. Therefore, designing robust algorithms with noisy labels is of great significance \cite{wei2019harnessing}.

In the literature, a lot of approaches were proposed, which can generally be classified into three categories: estimating the noise transition matrix \cite{goldberger2016training,hendrycks2018using,patrini2017making}, designing noise-robust loss functions \cite{ghosh2017robust,zhang2018generalized,xu2019l_dmi}, and sample correcting/selecting \cite{reed2014training,veit2017learning,tanaka2018joint,han2019deep,han2018co,chen2019understanding}. The schemes based on transition matrix estimating try to capture the transition probability between the noisy label and true label \cite{han2019deep}. 
Different transition matrix estimation methods were proposed in work \cite{goldberger2016training,hendrycks2018using,patrini2017making}, such as using additional softmax layer \cite{goldberger2016training}, utilizing trusted samples in a data-efficient manner \cite{hendrycks2018using}, and two-step estimating scheme \cite{patrini2017making}. However, these transition matrix estimations fail in real-world datasets where the utilized prior assumption is no longer valid \cite{han2019deep}. Being free of transition matrix estimation, the second category targets designing loss functions that have more noise-tolerant power. Work \cite{ghosh2017robust} adopted mean absolute error (MAE) function which demonstrates more noise robust ability than cross-entropy loss. In work \cite{zhang2018generalized}, the authors combined Generalized Cross Entropy (GCE) loss and MAE to address the slow convergence speed of work \cite{ghosh2017robust}. Recently, the authors in work \cite{xu2019l_dmi} proposed determinant-based mutual information loss which can be applied to any existing classification neural networks regardless of the noise pattern. However, this kind of method suffers generalization performance loss due to the low quality of the validation sets \cite{sun2019limited}.

The third category targets correcting noisy labels or selecting the possibly clean samples. Bootstrap in work \cite{reed2014training} adopted the predicted correcting labels together with the raw labels to lower the interference from the noisy samples. In work \cite{veit2017learning}, the authors utilized some clean annotations to reduce the noise in the large dataset. However, it is difficult to obtain the required certain amount of clean data in some cases. In work \cite{tanaka2018joint}, a joint optimization framework was proposed to gradually estimate the true labels. The self-learning framework was applied to train the label correction network without extra supervision \cite{han2019deep}. Some works just selected the clean data by dropping the noisy data directly to avoid the estimation of true labels. Co-teaching has appeared in the literature recently \cite{han2018co,chen2019understanding}, which trains two deep neural networks simultaneously and makes them teach each other by selecting some data of possibly clean labels. Compared with the first two types of methods, the third category is more general and can be integrated into many image classification tasks. 

Although many studies are proposed to suppress the noisy labels for the general image classification problem, there are few works on the classification of noised histopathology images. Work \cite{cheng2019robust} proposed online uncertainty sample mining and individual re-weighting methods to train their network. In work\cite{cao2020breast}, a double-softmax classification module was adopted to prevent overfitting the noisy labels and a teacher-student module was used to strengthen the effect of clean labels. Unfortunately, almost all these approaches fail in distinguishing informative hard samples from harmful mislabeled ones. Although the work in\cite{cheng2019robust} realized the significance of hard samples, their algorithm still didn't separate hard samples from noisy ones, and thus many important hard samples were mistakenly discarded \cite{xiao2015learning}.

On the one hand, the hard samples can make training more effective and efficient \cite{shrivastava2016training}. On the other hand, the deep neural networks can easily overfit to some label noise, and thus cause performance degradation \cite{li2019learning}. How to involve hard samples for training while reducing noise interference at the same time is of great significance. Many works found that the deep models first memorize the easy training data with clean labels and then memorize the hard or noisy data \cite{zhang2016understanding,wang2019symmetric}. This phenomenon can be used to distinguish the easy clean data from the noise, however how to distinguish the hard clean data from the noise is still not clear. 

In this work, we strive to reconcile this gap by proposing a hard sample aware noise robust learning algorithm. Our analysis reveals that the prediction history for each sample can be used as guiding information for distinguishing the hard and noisy samples. A deep model for hard and noisy sample detection is thus designed and integrated into our noisy label correction architecture. In the architecture, self-training is applied to conduct the label correction automatically. Based on the corrected data, the noise suppressing and hard enhancing (NSHE) scheme is designed to further enhance the hard sample and weaken the possible noisy sample. Our key contributions are summarized as follows. 

\begin{itemize}
\item We proposed a two-phase hard sample aware noise robust learning algorithm for histopathology image classification. Our method can save more clean samples by detecting the hard sample and noise in label correction phase. The hard samples can be further enhanced in our NSHE phase.

\item We built an EHN (easy/hard/noisy) detection scheme and integrated it into our self-learning label correction flow. We found that hard and noisy samples can be recognized using sample prediction history. Different from the previous works, our scheme can save more hard samples and discard more noisy samples.

\item In the NSHE phase, we smoothly trained our model to further suppress the interference from noisy samples and enhance the hard samples based on our proposed co-learning architecture. 

\item Our proposed method can be directly applied to the real-world dataset without using clean annotations. The experimental results verify that the proposed algorithm can achieve superior performance in our collected clinical pathology data of one top hospital in Beijing, China.

\end{itemize}


\section{Related Works}
\label{sec:RelatedWork}

\subsection{Noisy Label Correction}
Noisy label correction aims at improving the quality of the raw data by replacing the noisy labels with their true labels. Generally, the true labels are predicted by an extra model that is trained on a subset of clean data, such as work \cite{lee2018cleannet} and \cite{veit2017learning}. However, for the real-world dataset, the required clean dataset is not available and thus these methods will fail in this case.

To get rid of the dependence on clean samples, several noisy label correction methods \cite{han2019deep,tanaka2018joint} were proposed based on self-learning by pseudo-labels. In fact, pseudo-labeling is a type of self-training which is often used in semi-supervised learning scenario \cite{lee2013pseudo} with many unlabeled data \cite{tanaka2018joint}. In the semi-supervised learning scenario, the pseudo-labels are initially assigned to unlabeled data by predictions of a model that is trained on a labeled subset. This process is repeated and pseudo-labels are thus updated gradually. However, when processing the noisy datasets with labels, the challenge comes from the uncertainty about what is correct and what is incorrect in the data. In work \cite{tanaka2018joint}, the authors replaced all the labels with pseudo-labels to improve the quality of the original noisy dataset. The authors in work \cite{han2019deep} proposed a self-learning with multi-prototypes (SMP) scheme to train a robust model on the real-world noisy data.  

In the above self-learning based methods, all the pseudo labels are involved in the correction model training. However, misleading samples are inevitable, which will thus ruin the performance of the obtained correction model. To address this problem, we proposed an EHN detection scheme based on the prediction history of each sample to recognize the possible noisy samples. Then, we will remove these noisy samples in the correction model training and thus improve the label correction quality. Note that the output of EHN is also used in guiding the noisy sample discarding module (post-processing), which can save more hard samples for our NSHE scheme.

\subsection{Learning to Teach}
Learning to teach refers to the schemes that consist of two networks, the teacher and student networks \cite{gong2016teaching,fan2018learning}. The teacher tries to choose more informative samples to guide the training of student networks. However, these methods cannot process the dataset with noisy labels.

To make the above learning to teach algorithms be capable of processing noisy data, the authors in work \cite{jiang2017mentornet} proposed a novel MentorNet to supervise the training of the student network by focusing on the probably correct samples. However, the designed MentorNet suffers the disadvantage of accumulating error introduced by the sample-selection bias \cite{han2018co}. Another method called Decoupling proposed by work \cite{malach2017decoupling} trains two models simultaneously and updates the models by sampling with different predictions. However, in the selected subset with disagreement labels, there are still some noisy ones, which will decrease the performance of the trained model \cite{han2018co}. To solve these problems, based on Co-training of work \cite{blum1998combining}, the authors in \cite{han2018co} proposed a learning scheme called Co-teaching, which can train the model successfully even in the extremely noised dataset. Co-teaching also includes two models, and each model adopts the samples with small losses to train its peer network. Through the prediction information exchange, the error flows can be reduced accordingly. The authors in \cite{chen2019understanding} tried to improve the performance by proposing an Iterative Noisy Cross-Validation (INCV) method with the seriously noised dataset.


The above Co-teaching based methods try to conduct noise-robust learning by selecting clean samples with a small loss as much as possible. However, both hard and noisy samples have a large loss, and this will inevitably ignore the informative hard samples. We attempt to solve the problem in two aspects. First, we pre-discard most of the noisy samples with our hard-sample aware post-processing module. Second, we enhance most of the hard samples while suppressing the few existing noisy ones at the same time with our NSHE scheme.

\section{Method}
\label{sec:method}
\subsection{Architecture}

\begin{figure}[hbt]
  \centering
  \includegraphics[width=3.5in]{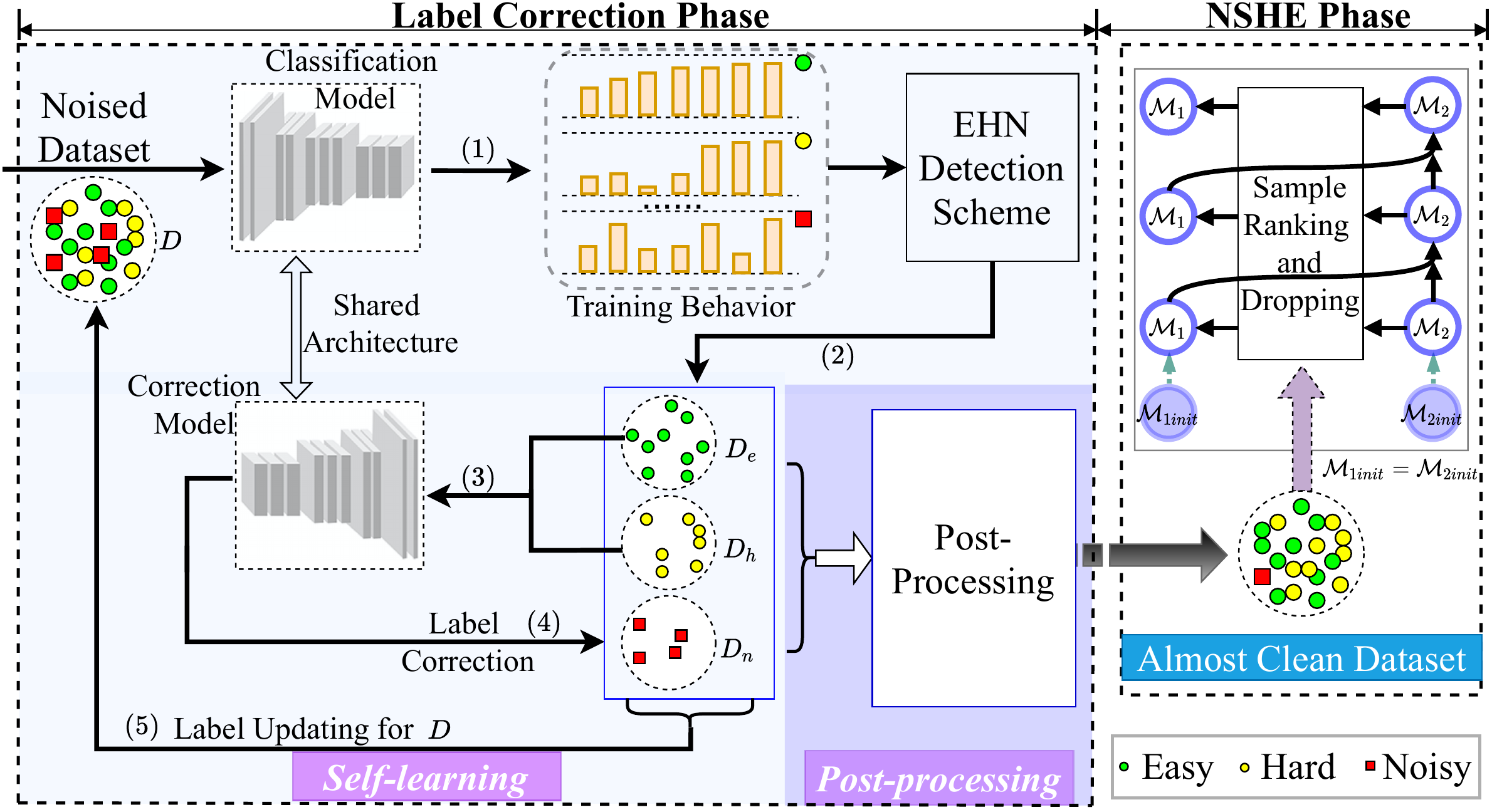}
  \caption{{The architecture of our algorithm. The architecture contains label correction phase and the NSHE phase. The label correction phase first generates ``Almost Clean Dataset'' and then puts it into our NSHE scheme. In the label correction phase, our self-learning contains (1) training history generation, (2) easy, hard, and noisy sample detection, (3) correction model training, (4) label correction, and (5) label updating. The post-processing is conducted after the self-learning flow. In the NSHE phase, two models are trained simultaneously with the obtained ``Almost Clean Dataset'' to produce the final robust classification model. First, $M_2$ selects training data for $M_1$. Then, $M_1$ and $M_2$ are updated by back-propagation and momentum manner, respectively.}}
  \label{fig:architecture}
\end{figure}

The top architecture contains two main phases: 1) label correction, 2) NSHE scheme, as depicted by Fig. \ref{fig:architecture}. The label correction takes the noisy data as the input and generates almost clean data. The NSHE scheme takes the corrected almost-clean dataset as the input and produces the final robust classification model. 

For the label correction, we designed a hard sample aware self-learning to achieve high-quality pseudo-labels and further cleaned our dataset by post-processing to drop out the possibly noisy samples for NSHE. The target of the label correction phase is to restore as many clean samples as possible. After performing label correction, the NSHE scheme aims at further reducing the impact of noisy and emphasizing the hard samples at the same time. Specifically, we smoothly trained our model based on the proposed co-learning architecture.

\subsection{Label Correction}
The proposed label correction architecture mainly consists of the classification/correction model, easy/hard/noisy (EHN) detection scheme, and post-processing component.

Firstly, the classification model is trained based on noisy data. By using this model, the prediction behavior for all the training samples can be obtained. Note that the prediction behavior means the prediction history for one sample through all the {$k$} (such as 30) training epochs. Then, the EHN detection scheme is applied to divide the dataset into three parts: easy, hard, and noisy. With the obtained easy and hard samples, the correction model is trained and used to correct the noisy data. Through repeating the above flow, the dataset quality is thus improved gradually. Finally, the dataset is further filtered by getting through the hard-aware post-processing component. In the following section, we will focus on the details of the EHN detection scheme, correction model, and post-processing component.

\textbf{EHN Detection Scheme.} 
Previous work \cite{zhang2016understanding} showed that CNNs tend to memorize simple samples first, and then the networks can gradually learn all the remaining samples, even including the noisy samples, due to the high representation capacity. However, overfit to the noise leads to poor generalization performance. To avoid the memorization of noisy data, work \cite{jiang2017mentornet} selected the samples with small loss to train the model, where such samples are treated as clean ones.

\begin{figure}[]
  \centering
  \includegraphics[width=3.49in]{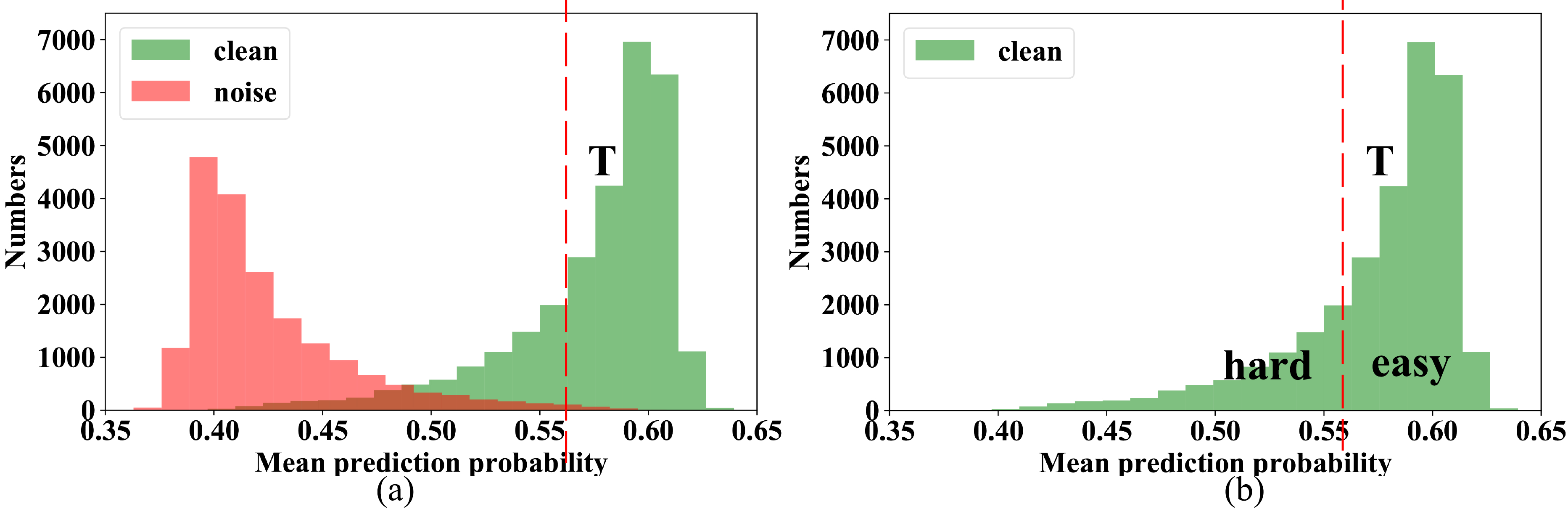}
  \caption{{(a) Mean prediction probability histogram of the clean and noisy samples in 40\% noise ratio dataset. (b) The division of easy and hard samples in clean.}}
  \label{fig:meanHistorgam}
\end{figure}

Sample with small loss means the prediction probability of the model output is closer to the supervising label. However, the normalized probability is much easier to analyze than the loss value. Different from work \cite{jiang2017mentornet}, we apply the mean prediction probability value of the sample training history in our EHN detection scheme. Fig. \ref{fig:meanHistorgam} shows the mean prediction probability histogram of clean and noisy samples. The figure shows most of the clean samples have higher mean prediction probabilities than the noisy ones. Therefore, we can set a threshold (such as the red dotted line in Fig. \ref{fig:meanHistorgam}) to preliminarily extract some clean samples, and we call these clean samples bigger than the threshold as easy samples. However, there's still a part of clean samples that's behind the threshold, and we can't distinguish them from noisy ones. We define this part of clean data as hard samples in our work. So far as we know, there are no existing schemes that can distinguish the hard from the noisy samples. 

\begin{figure*}[hbt]
  \centering
  \includegraphics[width=7.0in]{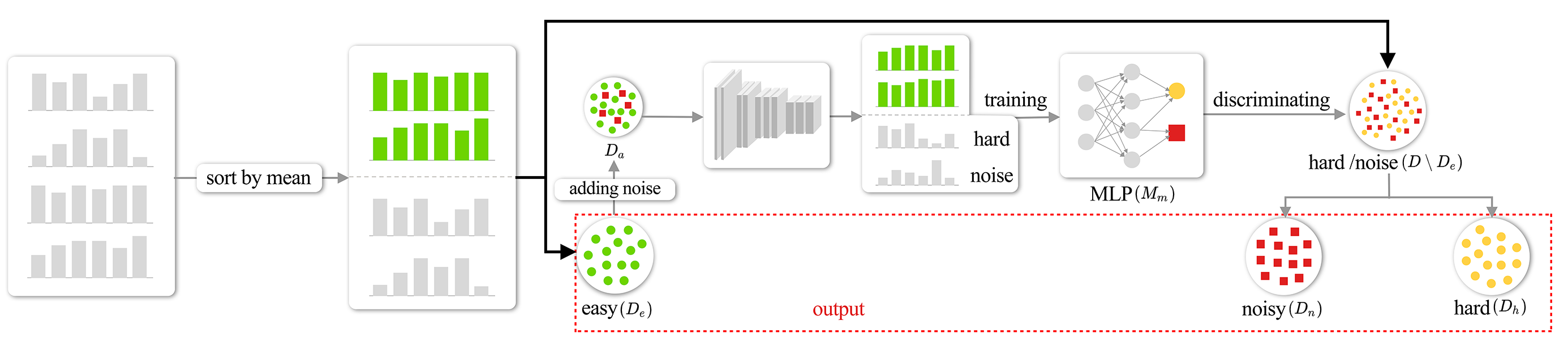}
  \caption{{The architecture of EHN detection scheme. First, the easy samples are selected by the mean prediction probability of training history. Second, adding noise to $D_e$ as $D_a$ and train a classification model on $D_a$ to get the training history. Third, $M_m$ is trained by the training history. At last, the hard and noisy samples in $D$ are distinguished by $M_m$.}}
  \label{fig:EHN-scheme}
\end{figure*}



We constructed our EHN detection scheme based on the prediction history of the training samples, as depicted by Fig. \ref{fig:EHN-scheme}. For the training set $D$ with $N$ samples, we gradually obtain the corresponding $N$ prediction probability maps through the training of a CNN classification model for {$k$} epochs. {Our EHN detection scheme first selected easy samples $D_e$ by using the mean prediction probabilities according to the threshold $T$. For convenience, the threshold $T$ is implemented as selected easy sample ratio $\tau_e$ in this paper. The higher $\tau_e$ corresponds to the lower threshold $T$, and vice versa. Then we added noise to the $D_e$ as $D_a$ by switching the labels of some samples in $D_e$, and recorded whether the samples are noise or not as $R$}. The noise ratio of the adding noise is the same as the original dataset, which can be estimated by the noise cross-validation algorithm of \cite{chen2019understanding}. After that, we trained the same classification model by using $D_a$ and recorded training history again. Then we discarded the ``easy samples'' of $D_a$ according to mean prediction probabilities and utilized the rest samples as training data for the Multi-Layer Perceptron (MLP) classifier. So far, we will obtain an MLP classifier that takes prediction probability maps of training history as input and output whether it is a hard sample or a noisy one. Finally, we put the samples in $D \setminus D_e$ into the MLP classifier and get the hard sample set $D_h$ and the noisy set $D_n$. Algorithm \ref{alg:ehn} shows the details of the EHN detection scheme.

\begin{algorithm}
	\renewcommand{\algorithmicrequire}{\textbf{Input:}}
	\renewcommand{\algorithmicensure}{\textbf{Output:}}
	\caption{{EHN detection scheme}}
	\label{alg:ehn}
	\begin{algorithmic}[1]
		\REQUIRE  $D=[d_1,d_2,...,{d_N}]$, $d_i$ is input image, label $Y=[y_1,y_2,...,{y_N}]$, $y_i$ is label for $d_i$, easy sample ratio $\tau_e$
		\ENSURE easy set $D_e$, hard set $D_h$, noisy set $D_n$
		\STATE Train classification model $M_c$ by using $D$ and $Y$, record training history $H=[h_1,h_2,...,{h_N}]$, where $h_i$ is a vector with shape of $1 * {k}$ (epoch)
		\STATE Calculate the mean value of $H$ as $H_m$, $H_m = [mean(h_1),mean(h_2),...,mean({h_N})]$, sort $D$ descending by $H_m$, select easy samples $D_e = D[0:len(D) * \tau_e]$
        \STATE Add noise to $D_e$ as $D_a$, get noisy labels {$Y_a$} and record whether the samples are noise or not $R=[r_1,r_2,...,{r_N}]$
        \STATE Retrain $M_c$ by $D_a$ and {$Y_a$}, record training history {${H_a}$}
        \STATE Sort {$H_a$} descending by mean value, select training history of hard and noisy samples {$H_a’ = H_a[len(H_a)*\tau_e:len(H_a)]$}, and choose the corresponding $R'$ from $R$
		\STATE Train MLP classifier $M_m$ by using {$H_a'$} and $R'$ 
        \STATE Put the samples in $D \setminus D_e$ into $M_m$ and get $D_h$ and $D_n$
		\STATE \textbf{return} $D_e$, $D_h$, $D_n$
	\end{algorithmic}  
\end{algorithm}

\begin{algorithm}
	\renewcommand{\algorithmicrequire}{\textbf{Input:}}
	\renewcommand{\algorithmicensure}{\textbf{Output:}}
	\caption{Post-processing strategy}
	\label{alg:post-p}
	\begin{algorithmic}[1]
		\REQUIRE easy set $D_e$, hard set $D_h$, noisy set $D_n$, original labels $Y$, pseudo labels $G$ generated from the correction model
		\ENSURE almost clean dataset $D_o$
		\STATE $y_i$ = label of sample $d_i$ in $D_h \cup D_n$,  in $Y$
		\STATE $g_i$ = pseudo label of sample $d_i$ in $D_h \cup D_n$,  in $G$
		\STATE Initialize $N$ = $\varnothing$ as noisy sample set
		
		\FORALL{$d_i \in D_h \cup D_n$}
        \IF{($y_i$ = $g_i$ and $d_i \in D_n$) or ($y_i$ $\neq$ $g_i$ and  $d_i \in D_h$)} \STATE{$N$  = $N\cup \lbrace d_i\rbrace$} 
        \ELSE {} \STATE{$y_i$ = $g_i$}
        \ENDIF
		\ENDFOR
        \STATE $D_o$  = $(D_e \cup D_h \cup D_n) \setminus N$ 
		\STATE \textbf{return} $D_o$
	\end{algorithmic}  
\end{algorithm}

\textbf{Label Correction Model.} Our correction model is trained by using $D_e \cup D_h$ from the EHN detection scheme. After training, the model has some ability to correct noisy labels. Therefore, the labels of samples in $D_h \cup D_n$ are replaced by the pseudo labels generated from the correction model, where the pseudo labels are the class with the highest probability of model output. The reason we also put the hard samples into the correction model is that we cannot trust the result of the MLP classifier in our EHN detection scheme completely.

These steps above are called our self-learning flow; it takes the original dataset into the classification model and gets through the EHN detection scheme by using Algorithm \ref{alg:ehn}. Then, it trains the correction model and updates some sample labels with the pseudo ones. Finally, it iterates over the above steps to further purify our dataset.

\textbf{Post-processing Component.} Our post-processing component is to drop out the noisy samples which still can not be corrected after the processing of EHN and label correction. {In $D_n$, we drop out the samples whose labels were not changed by the correction model. In $D_h$, we drop out the samples whose labels were changed by the correction model.} Algorithm \ref{alg:post-p} shows the details of the post-processing component. After post-processing, we then obtain the almost-clean dataset.

\subsection{Noise Suppressing and Hard Enhancing (NSHE)}

\begin{figure}[hbt]
  \centering
  \includegraphics[width=1.4in]{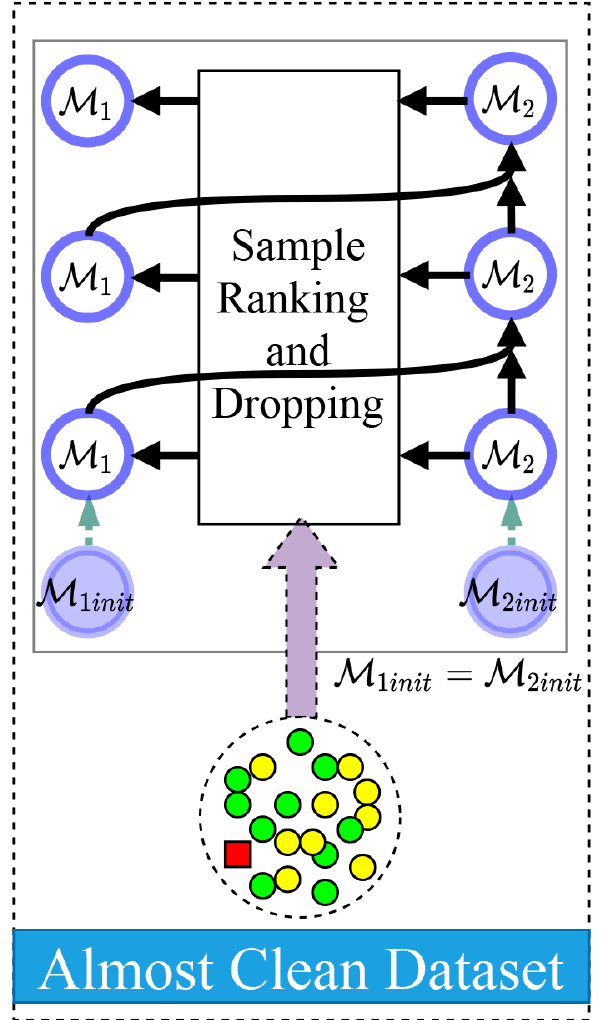}
  \caption{{The NSHE phase takes the obtained ``Almost Clean Dataset'' as input and output the final noise-robust classification model. For the training process, it initializes two models $M_1$ and $M_2$ with the same parameters. At each epoch, $M_2$ selects training data for $M_1$ by ranking the prediction values of samples. $M_1$ is updated by back-propagation, and $M_2$ is updated by $M_1$ and the previous $M_2$.}}
  \label{fig:NSHE}
\end{figure}

\begin{algorithm}
	\renewcommand{\algorithmicrequire}{\textbf{Input:}}
	\renewcommand{\algorithmicensure}{\textbf{Output:}}
	\caption{{NSHE scheme}}
	\label{alg:alg3}
	\begin{algorithmic}[1]
		\REQUIRE almost-clean data set $D_o$, label $Y$, noise discarding ratio $\tau$, momentum ratio $m$
        \ENSURE classification model $M$
        \STATE Initialize $\theta_1$ of $M_1$, $\theta_2$ of $M_2$, let $\theta_1 = \theta_2$
		\FOR{$epoch = 1,2,...., MaxEpoch$}
            \STATE get prediction probabilities $P_2$ = $M_2(D_o,Y)$, sort $D_o$ ascending by $P_2$
            \STATE obtain discarded sample set $N$ = $D_o[0:len(D_o)*\tau]$
        	\FOR{$iter = 1,2,...., MaxIter$}
            	\STATE sample mini-batch $B$
            	\STATE $B = B \setminus N$
                \STATE get prediction probabilities $P_1$ = $M_1(B,Y)$
                \STATE $L = FL(P_1)$
                \STATE $\theta_1 = SGD(L,\theta_1)$
                \STATE $\theta_{\mathrm{2}} = m \theta_{\mathrm{2}}+(1-m) \theta_{\mathrm{1}}$
        	\ENDFOR
		\ENDFOR
        \STATE \textbf{return} $M_2$
	\end{algorithmic}  
\end{algorithm}

Here we developed our robust NSHE algorithm by using the almost-clean dataset. {The overview of the NHSE phase is shown in Fig. \ref{fig:NSHE}.} It is found in the experiment that different samples may have completely opposite optimization directions for model parameter values, which leads to frequent dithering of model parameters during the training process, resulting in a poor effect. This phenomenon is even more serious in noisy dataset, and the noisy samples will mislead the model training.  Inspired by MoCo\cite{he2020momentum},  we initialized two models {$M_1$, $M_2$} with the same backbone and parameters. Formally, denoting the parameters of {$M_1$} as $\theta_1$ and those of {$M_2$} as $\theta_2$, we update $\theta_2$ by:

\begin{equation}
\theta_{\mathrm{2}} \leftarrow m \theta_{\mathrm{2}}+(1-m) \theta_{\mathrm{1}}
\end{equation}
Here $m \in[0,1)$  is a momentum coefficient. Only the parameters $\theta_1$ are updated by back-propagation. The momentum update in (1) makes $\theta_2$ evolve more smoothly than $\theta_1$. Since the almost-clean dataset still has some noisy samples, we ranked the samples according to the prediction probabilities of the labeled class at each epoch, and set a very small ratio to make the samples with small prediction probabilities unable to participate in back-propagation. To avoid confirmation bias, we proposed the co-learning architecture based on Co-teaching\cite{han2018co}. The probabilities are computed by {$M_2$}, namely the sample selection information was given by {$M_2$}. To further emphasize the significance of hard samples, we used focal loss \cite{lin2017focal} to strengthen hard samples. The loss function is defined as follows:

\begin{equation}
\mathrm{FL}\left(p_{\mathrm{t}}\right)=-\left(1-p_{t}\right)^{\gamma} \log \left(p_{\mathrm{t}}\right)
\end{equation}
where $p_{t}$ is the predicted probability of the correct class, $\gamma$ is a hyper-parameter. Algorithm \ref{alg:alg3} shows the training flow.

\section{Experiment}
\label{sec:experiment}

\subsection{Dataset}
\label{sec:data}

\begin{figure}[hbt]
  \centering
  \includegraphics[width=3.8in]{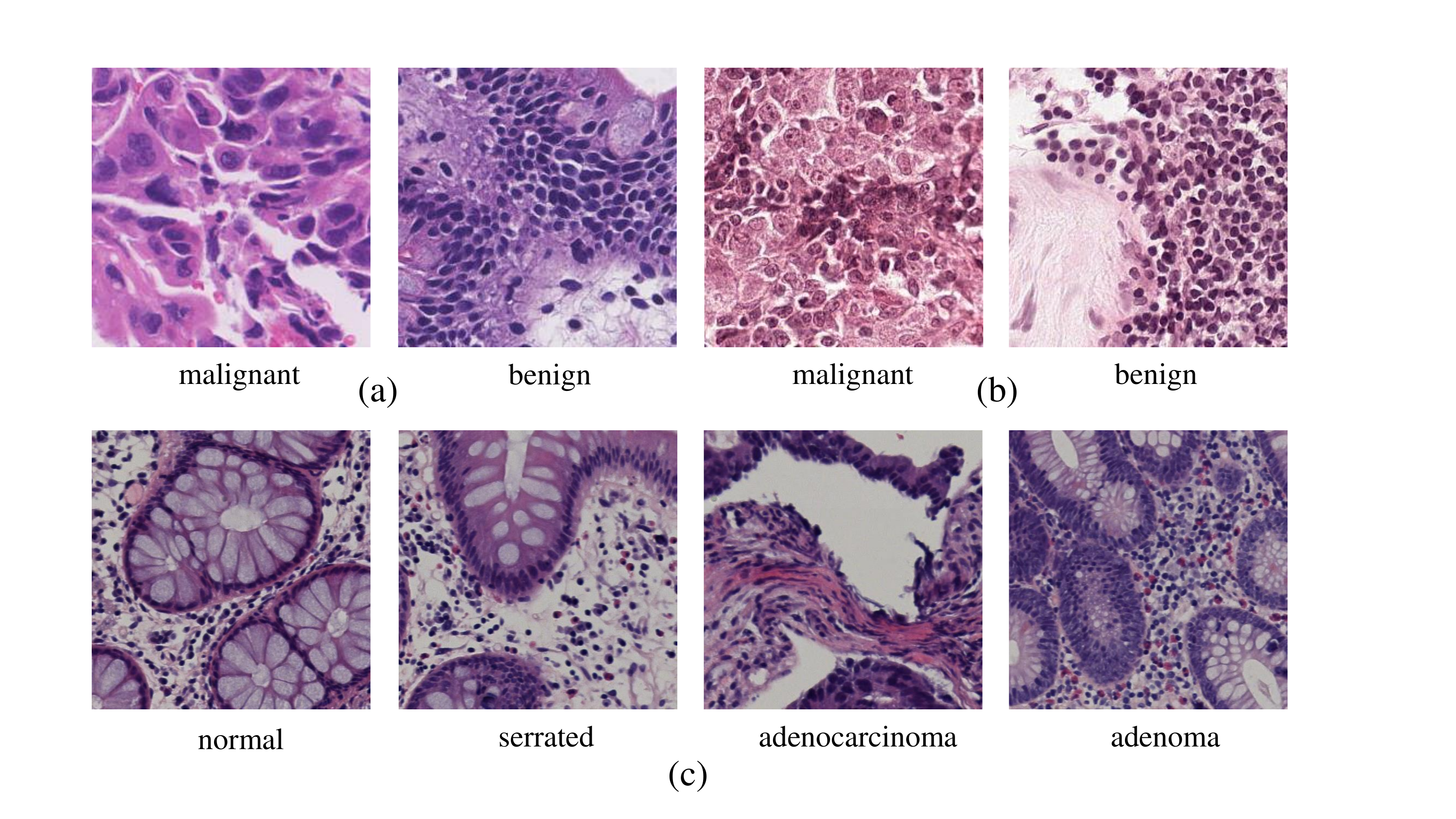}
  \caption{Selected samples. (a) DigestPath2019. (b) Camelyon16. (c) Chaoyang.}
  \label{fig:dataset}
\end{figure}

We extensively validate our method on {five} datasets. 
\begin{enumerate}
\item DigestPath2019: It has 250 malignant images with pixel-level annotation and 410 benign images. We cropped all images into small patches, using a patch size of $256 \times 256$ and a stride of 64 pixels. {With the segmentation annotations, we defined patches with the lesion area accounts for more than $95\%$ of the whole patch as malignant samples. All benign samples were cropped from benign images.} Finally, we got 29334 malignant samples, 28419 benign samples, and randomly partition them into 24611 malignant, 23824 benign, and 4723 malignant, 4595 benign samples for training and testing, respectively. The training and testing patches were from different original images. The sample patches are shown in Fig. \ref{fig:dataset} (a).
\item Camelyon16: It has 110 tumor WSIs (whole slide images) and 110 normal WSIs, and we preprocessed it in the same way as the DigestPath2019 dataset. Finally, we got 16050 malignant samples, 14812 negative samples, and randomly partition them into 11262 malignant, 11052 benign, and 4788 malignant, 4760 benign samples for training and testing, respectively. Similarly, the training and testing patches were from different WSIs. The sample patches are shown in Fig. \ref{fig:dataset} (b).
\item Chaoyang: Colon slides from Chaoyang hospital, the patch size is $512 \times 512$. We invited 3 professional pathologists to label the patches, respectively. We took the parts of labeled patches with consensus results from 3 pathologists as the testing set. Others we used as the training set. For the samples with inconsistent labeling opinions of the three doctors in the training set {(this part accounts for about 40\%)}, we randomly selected the opinions from one of the three doctors. Finally, we got 1111 normal, 842 serrated, 1404 adenocarcinoma, 664 adenoma, and 705 normal, 321 serrated, 840 adenocarcinoma, 273 adenoma samples for training and testing, respectively. This noisy dataset is constructed in the real scenario. Fig. \ref{fig:dataset} (c) shows the sample patches.
{
\item CIFAR-10 \cite{2009Learning}: It consists of 60000 colour images with size $32 \times 32$ in 10 classes, with 6000 images per class. There are 50000 training images and 10000 test images. The classes in the dataset are airplane, automobile, bird, cat, deer, dog, frog, horse, ship, and trunk. It is a popular public natural computer vision dataset for image classification.
}
{\item Webvision \cite{2017WebVision}: It contains 2.4 million images in 1000 classes, which crawled from websites. The training set contains many real-world noisy labels. Since the dataset is quite large, for quick experiments, we follow the previous work\cite{chen2019understanding} and only use the first 50 classes of the Google image subset. Finally, it contains 65944 samples for training and 2500 samples for testing.
}
\end{enumerate}

{Among them, 1) to 3) are medical scenario datasets; 4) and 5) are natural computer vision datasets. We randomly added different ratios (10\%, 20\%, 30\%, and 40\%) of noise to the DigestPath2019 and the Camelyon16. Due to these two datasets only have two classes, the noise type is simply changing labels into another class. For CIFAR-10 dataset, as it originally does not contain label noise, following previous work \cite{chen2019understanding}, we experiment with two types of label noise: symmetric and asymmetric. Symmetric noise is generated by randomly replacing the labels for a percentage of the training data with all other classes, and asymmetric noise is only generated by replacing the labels with adjacent class. Following work \cite{chen2019understanding}, we tested noise ratios 20\%, 50\%, and 80\% for symmetric noise, and noise ratio 40\% for asymmetric noise. The Chaoyang and Webvision datasets are constructed in the real scenario, and the noise refers to the actual labeled samples that are wrong, rather than the artificial addition.}

\begin{table*}[]
\scriptsize
\centering
\caption{{Average test ACC, F1 Score, AUC, {Precision}, Recall(\%, 3 runs) with standard deviation on DigestPath2019 dataset.}}
\label{table:objMiccai}
\begin{tabular}{@{}lllllllllll@{}}
\toprule
\multicolumn{1}{c}{Noise ratio} & \multicolumn{5}{c}{10\%}                                                                                                                & \multicolumn{5}{c}{20\%}                                                                                                                \\ \midrule
Method                & \multicolumn{1}{c}{ACC} & \multicolumn{1}{c}{AUC} & \multicolumn{1}{c}{F1 Score} & \multicolumn{1}{c}{{Precision}} & \multicolumn{1}{c}{Recall} & \multicolumn{1}{c}{ACC} & \multicolumn{1}{c}{AUC} & \multicolumn{1}{c}{F1 Score} & \multicolumn{1}{c}{{Precision}} & \multicolumn{1}{c}{Recall} \\ \midrule
Joint                     & 91.49$\pm$0.95          & 97.62$\pm$1.25          & 91.25$\pm$0.90               & 95.41$\pm$1.77          & 87.45$\pm$0.21             & 88.78$\pm$0.28          & 95.40$\pm$0.32          & 89.00$\pm$0.27               & 88.47$\pm$0.24          & 89.53$\pm$0.32             \\
Co-tea.                     & 91.08$\pm$0.79          & 98.21$\pm$1.00          & 90.59$\pm$0.80               & \textbf{97.49$\pm$1.18} & 84.60$\pm$0.51             & 90.33$\pm$0.70          & 95.86$\pm$0.87          & 90.12$\pm$0.65               & 93.49$\pm$1.29          & 87.00$\pm$0.11             \\
INCV                            & 93.58$\pm$0.96          & 97.91$\pm$1.01          & 93.62$\pm$0.96               & 94.34$\pm$0.92          & 92.92$\pm$1.00             & 92.09$\pm$0.39          & 96.43$\pm$0.50          & 91.97$\pm$0.38               & \textbf{94.74$\pm$0.65} & 89.36$\pm$0.14             \\
OUSM                       & 90.27$\pm$0.98          & 96.87$\pm$1.02          & 89.81$\pm$0.94               & 95.87$\pm$1.95          & 84.48$\pm$0.16             & 88.19$\pm$1.56          & 94.19$\pm$1.48          & 88.18$\pm$1.69               & 89.30$\pm$0.85          & 87.11$\pm$2.49             \\
NF-Net                          & 83.08$\pm$0.43          & 90.64$\pm$0.24          & 81.95$\pm$0.42               & 89.26$\pm$0.73          & 75.75$\pm$0.18             & 83.09$\pm$2.17          & 89.09$\pm$2.45          & 82.82$\pm$2.21               & 85.40$\pm$2.28          & 80.40$\pm$2.15             \\
DM                       & 92.69$\pm$0.84          & 97.76$\pm$0.64          & 92.69$\pm$0.81               & 93.98$\pm$1.30          & 91.45$\pm$0.59             & 90.53$\pm$0.87          & 96.09$\pm$1.11          & 90.46$\pm$0.83               & 92.51$\pm$1.49          & 88.51$\pm$0.96             \\
SELF                            & 92.22$\pm$0.77          & 97.43$\pm$0.96          & 92.21$\pm$0.79               & 92.55$\pm$1.04          & 91.78$\pm$0.66             & 91.23$\pm$0.42          & 95.89$\pm$0.55          & 91.19$\pm$0.36               & 91.44$\pm$0.33          & 90.99$\pm$0.41             \\
\textbf{Ours}                   & \textbf{94.91$\pm$0.32} & \textbf{98.40$\pm$0.57} & \textbf{95.05$\pm$0.35}      & 93.68$\pm$0.63          & \textbf{96.46$\pm$0.21}    & \textbf{94.46$\pm$0.20} & \textbf{98.30$\pm$0.34} & \textbf{94.53$\pm$0.23}      & 94.51$\pm$0.13          & \textbf{94.55$\pm$0.38}    \\ \midrule
\multicolumn{1}{c}{Noise ratio} & \multicolumn{5}{c}{30\%}                                                                                                                & \multicolumn{5}{c}{40\%}                                                                                                                \\ \midrule
Method                & \multicolumn{1}{c}{ACC} & \multicolumn{1}{c}{AUC} & \multicolumn{1}{c}{F1 Score} & \multicolumn{1}{c}{{Precision}} & \multicolumn{1}{c}{Recall} & \multicolumn{1}{c}{ACC} & \multicolumn{1}{c}{AUC} & \multicolumn{1}{c}{F1 Score} & \multicolumn{1}{c}{{Precision}} & \multicolumn{1}{c}{Recall} \\ \midrule
Joint                     & 87.17$\pm$0.57          & 94.72$\pm$0.63          & 86.41$\pm$0.66               & 93.34$\pm$0.27          & 80.44$\pm$0.98             & 84.30$\pm$0.87          & 91.39$\pm$1.02          & 84.46$\pm$0.92               & 84.77$\pm$0.60          & 84.15$\pm$1.26             \\
Co-tea.                     & 86.49$\pm$0.37          & 93.55$\pm$0.46          & 86.90$\pm$0.33               & 85.50$\pm$0.47          & 88.34$\pm$0.20    & 84.92$\pm$0.57          & 92.98$\pm$0.68          & 84.94$\pm$0.54               & 86.02$\pm$0.74          & 83.89$\pm$0.41             \\
INCV                            & 87.84$\pm$0.52          & 94.60$\pm$0.23          & 87.65$\pm$0.52               & 90.36$\pm$0.56          & 85.10$\pm$0.49             & 85.20$\pm$0.48          & 93.52$\pm$0.56          & 84.47$\pm$0.50               & 90.19$\pm$0.56          & 79.44$\pm$0.48             \\
OUSM                       & 86.52$\pm$1.22          & 92.49$\pm$1.25          & 86.53$\pm$1.23               & 87.64$\pm$1.18          & 85.45$\pm$1.37             & 84.04$\pm$0.38          & 93.51$\pm$0.42          & 82.57$\pm$0.41               & 92.43$\pm$0.48          & 74.60$\pm$0.36             \\
NF-Net                          & 83.86$\pm$0.59          & 90.62$\pm$0.34          & 83.59$\pm$0.53               & 86.24$\pm$0.93          & 81.11$\pm$0.18             & 82.37$\pm$0.84          & 88.29$\pm$1.02          & 82.38$\pm$0.81               & 83.49$\pm$0.97          & 81.30$\pm$0.70             \\
DM                       & 88.87$\pm$1.03          & 95.10$\pm$0.78          & 88.60$\pm$1.12               & 92.04$\pm$0.54          & 85.41$\pm$1.63             & 85.96$\pm$0.37          & \textbf{96.87$\pm$0.38} & 84.49$\pm$0.43               & \textbf{95.97$\pm$0.39} & 75.46$\pm$0.46             \\
SELF                            & 90.03$\pm$1.21          & 95.78$\pm$0.64          & 89.96$\pm$1.42               & 90.97$\pm$0.89          & \textbf{88.66$\pm$1.66}             & 86.23$\pm$0.98          & 94.21$\pm$0.89          & 86.02$\pm$1.00               & 87.52$\pm$0.88          & 84.03$\pm$1.21             \\
\textbf{Ours}                   & \textbf{91.72$\pm$0.69} & \textbf{97.38$\pm$0.78} & \textbf{91.47$\pm$0.75}      & \textbf{95.74$\pm$0.34} & 87.57$\pm$1.10             & \textbf{87.15$\pm$0.41} & 94.26$\pm$0.46          & \textbf{86.49$\pm$0.41}      & 88.58$\pm$0.63          & \textbf{85.16$\pm$0.25}    \\ \bottomrule
\end{tabular}
\end{table*}

\begin{table*}[]
\scriptsize
\centering
\caption{{Average test ACC, F1 Score, AUC, {Precision}, Recall(\%, 3 runs) with standard deviation on Camelyon16 dataset.}}
\label{table:objCamelyon}
\begin{tabular}{@{}lllllllllll@{}}
\toprule
\multicolumn{1}{c}{Noise ratio} & \multicolumn{5}{c}{10\%}                                                                                                                & \multicolumn{5}{c}{20\%}                                                                                                                \\ \midrule
Method                & \multicolumn{1}{c}{ACC} & \multicolumn{1}{c}{AUC} & \multicolumn{1}{c}{F1 Score} & \multicolumn{1}{c}{{Precision}} & \multicolumn{1}{c}{Recall} & \multicolumn{1}{c}{ACC} & \multicolumn{1}{c}{AUC} & \multicolumn{1}{c}{F1 Score} & \multicolumn{1}{c}{{Precision}} & \multicolumn{1}{c}{Recall} \\ \midrule
Joint                     & 97.51$\pm$0.56          & 98.94$\pm$0.67          & 97.53$\pm$0.55               & 96.89$\pm$0.68          & 98.18$\pm$0.42             & 96.51$\pm$0.70          & 98.75$\pm$0.68          & 96.49$\pm$0.72               & 97.13$\pm$0.15          & 95.86$\pm$1.32             \\
Co-tea.                     & 98.16$\pm$0.30          & 99.34$\pm$0.27          & 98.13$\pm$0.31               & \textbf{99.87$\pm$0.09} & 96.45$\pm$0.51             & 97.57$\pm$0.27          & 99.29$\pm$0.30          & 97.53$\pm$0.27               & 99.29$\pm$0.27          & 95.84$\pm$0.27             \\
INCV                            & 97.87$\pm$0.16          & 99.23$\pm$0.16          & 97.84$\pm$0.16               & 99.57$\pm$0.18          & 96.17$\pm$0.15             & 97.65$\pm$0.12          & 99.33$\pm$0.13          & 97.62$\pm$0.12               & 99.42$\pm$0.07          & 95.87$\pm$0.17             \\
OUSM                       & 98.08$\pm$0.41          & 99.37$\pm$0.50          & 98.06$\pm$0.41               & 99.55$\pm$0.55          & 96.61$\pm$0.52             & 96.57$\pm$1.11          & 98.90$\pm$0.20          & 96.52$\pm$1.13               & 98.33$\pm$1.21          & 94.78$\pm$1.33             \\
NF-Net                          & 90.24$\pm$0.21          & 93.53$\pm$0.19          & 89.70$\pm$0.24               & 95.16$\pm$0.05          & 84.84$\pm$0.39             & 90.72$\pm$1.02          & 92.93$\pm$0.45          & 90.30$\pm$1.03               & 94.92$\pm$1.55          & 86.12$\pm$0.68             \\
DM                       & 95.10$\pm$0.40          & 96.70$\pm$0.50          & 94.91$\pm$0.43               & 99.15$\pm$0.19          & 91.01$\pm$0.63             & 93.98$\pm$0.63          & 96.90$\pm$0.62          & 94.01$\pm$0.66               & 93.85$\pm$0.34          & 94.17$\pm$1.10             \\
SELF                            & 97.89$\pm$0.42          & 99.02$\pm$0.34          & 97.85$\pm$0.39               & 98.02$\pm$0.52          & 97.53$\pm$0.39             & 97.44$\pm$0.32          & 98.89$\pm$0.15          & 97.49$\pm$0.41               & 97.88$\pm$0.50          & 97.06$\pm$0.37             \\
\textbf{Ours}                   & \textbf{98.82$\pm$0.20} & \textbf{99.81$\pm$0.14} & \textbf{98.81$\pm$0.20}      & 99.73$\pm$0.18          & \textbf{97.73$\pm$0.22}    & \textbf{98.61$\pm$0.11} & \textbf{99.78$\pm$0.18} & \textbf{98.69$\pm$0.11}      & \textbf{99.61$\pm$0.14} & \textbf{97.40$\pm$0.21}    \\ \midrule
\multicolumn{1}{c}{Noise ratio} & \multicolumn{5}{c}{30\%}                                                                                                                & \multicolumn{5}{c}{40\%}                                                                                                                \\ \midrule
Method                & \multicolumn{1}{c}{ACC} & \multicolumn{1}{c}{AUC} & \multicolumn{1}{c}{F1 Score} & \multicolumn{1}{c}{{Precision}} & \multicolumn{1}{c}{Recall} & \multicolumn{1}{c}{ACC} & \multicolumn{1}{c}{AUC} & \multicolumn{1}{c}{F1 Score} & \multicolumn{1}{c}{{Precision}} & \multicolumn{1}{c}{Recall} \\ \midrule
Joint                     & 96.10$\pm$0.55          & 98.79$\pm$0.64          & 96.03$\pm$0.55               & 98.09$\pm$0.75          & 94.06$\pm$0.38             & 91.97$\pm$1.72          & 95.51$\pm$1.43          & 91.56$\pm$1.82               & 96.94$\pm$3.08          & 86.86$\pm$2.79             \\
Co-tea.                     & 97.49$\pm$0.12          & 99.02$\pm$0.11          & 97.44$\pm$0.12               & 99.68$\pm$0.04          & 95.30$\pm$0.20             & 95.20$\pm$0.71          & 96.25$\pm$0.72          & 95.09$\pm$0.73               & 97.56$\pm$0.77          & 92.74$\pm$0.69             \\
INCV                            & 96.28$\pm$0.32          & 97.64$\pm$0.32          & 96.24$\pm$0.31               & 97.41$\pm$0.43          & 95.10$\pm$0.20             & 95.40$\pm$0.23          & 96.26$\pm$0.34          & 95.28$\pm$0.24               & 98.06$\pm$0.10          & 92.65$\pm$0.36             \\
OUSM                       & 96.43$\pm$0.23          & 98.73$\pm$0.20          & 96.32$\pm$0.24               & 99.63$\pm$0.29          & 93.22$\pm$0.32             & 91.98$\pm$3.21          & 96.35$\pm$3.07          & 91.32$\pm$3.71               & 98.64$\pm$1.29          & 85.11$\pm$5.40             \\
NF-Net                          & 89.61$\pm$0.60          & 95.81$\pm$0.19          & 89.28$\pm$0.59               & 92.50$\pm$0.93          & 86.29$\pm$0.35             & 85.53$\pm$1.18          & 91.72$\pm$0.75          & 85.34$\pm$1.28               & 86.76$\pm$0.86          & 93.96$\pm$1.66             \\
DM                       & 93.39$\pm$1.07          & 95.28$\pm$1.19          & 93.08$\pm$1.17               & 97.74$\pm$0.60          & 88.85$\pm$1.64             & 89.55$\pm$0.12          & 97.72$\pm$0.06          & 88.39$\pm$0.14               & \textbf{99.84$\pm$0.12} & 79.30$\pm$0.15             \\
SELF                            & 96.98$\pm$0.54          & 98.72$\pm$0.17          & 97.00$\pm$0.46               & 97.56$\pm$0.38          & 96.43$\pm$0.56             & 96.02$\pm$1.05          & 97.53$\pm$0.31          & 95.89$\pm$1.22               & 96.78$\pm$1.31          & 94.95$\pm$1.20             \\
\textbf{Ours}                   & \textbf{98.32$\pm$0.41} & \textbf{99.57$\pm$0.22} & \textbf{98.30$\pm$0.42}      & \textbf{99.84$\pm$0.09} & \textbf{96.80$\pm$0.46}    & \textbf{98.17$\pm$0.84} & \textbf{99.51$\pm$0.24} & \textbf{98.16$\pm$0.28}      & 99.12$\pm$0.23          & \textbf{97.22$\pm$0.32}    \\ \bottomrule
\end{tabular}
\end{table*}

\subsection{Implementation and Parameter Settings}
\label{sec:impl}

For medical scenario datasets (DigestPath2019, Camelyon16, and Chaoyang). We used the Resnet-34 as the backbone and trained it using Adam with a momentum of 0.9, and a batch size of 96. During the label correction phase, the network was trained for 30 epochs. We set the initial learning rate as 0.001, and linearly reduced it after 15 epochs. For the NSHE phase, the networks were trained for 40 epochs. We set the initial learning rate as 0.001, and linearly reduced it after 15 epochs.

{For natural computer vision datasets (CIFAR-10, Webvision), we followed the same settings in work \cite{chen2019understanding}. For CIFAR-10, we used the Resnet-32 as the backbone and trained it using SGD with a momentum of 0.9, a learning rate of 0.02, and a batch size of 128. The networks were trained for 300 epochs both in the label correction phase and the NSHE phase. For Webvision, we used the Inception-Resnet v2 \cite{2017Inception} as the backbone and trained it using SGD with a momentum of 0.9, a learning rate of 0.01, and a batch size of 32. The networks were trained for 80 epochs both in the label correction phase and the NSHE phase.}

{For all the datasets, $\tau_e$ was set to 0.1 for 80\% noise ratio, and $1-1.5*\rho$ for other noise ratios, where $\rho$ is the dataset noise ratio. The parameter $\gamma$ in focal loss we set to $2$, and the discarding ratio $\tau$ was set to $0.1*\rho$. For real-world datasets Chaoyang and Webvision, $\rho$ was estimated by the noise cross-validation algorithm of \cite{chen2019understanding}.}

\subsection{Evaluation Criteria}

We used Accuracy (ACC), {Precision}, Recall, F1 Score (F1), AUC, and {ROC} curve as evaluation criteria. Their definitions are as follows:

\begin{equation}
ACC=\frac{TP + TN}{TP+TN+FP+FN}
\end{equation}

\begin{equation}
{Precision}=\frac{TP}{TP+FP}
\end{equation}

\begin{equation}
Recall=\frac{TP}{TP+FN}
\end{equation}

\begin{equation}
F1 =\frac{2*{Precision}*Recall}{{Precision}+Recall}
\end{equation}
where TP, TN, FP, and FN represent true positives, true negatives, false positives, and false negatives, respectively.

{ROC} curve is the receiver operating characteristic curve. Its abscissa is false positive rate and ordinate is the true positive rate. AUC is the area under the {ROC} curve. 

For multi-classification tasks, we compute the {Precision}, Recall, F1 Score, {ROC} curve, and AUC for each class and average them by using macro-average.
\subsection{Objective Comparison}
We compare our methods with the following methods using the same network architecture.

\begin{table}[]
\scriptsize
\centering
\caption{{Average test ACC, F1 Score, AUC, {Precision}, Recall(\%, 3 runs) with standard deviation on Chaoyang dataset.}}
\label{table:objChaoyang}
\begin{tabular}{@{}llllll@{}}
\toprule
Method & \multicolumn{1}{c}{ACC} & \multicolumn{1}{c}{AUC} & \multicolumn{1}{c}{F1 Score} & \multicolumn{1}{c}{{Precision}} & \multicolumn{1}{c}{Recall} \\ \midrule
Joint      & 75.99$\pm$0.64          & 90.43$\pm$0.84          & 67.72$\pm$2.36               & 70.97$\pm$2.42          & 67.91$\pm$2.77             \\
Co-tea.      & 79.39$\pm$0.29          & 91.72$\pm$0.68          & 71.97$\pm$0.96               & 74.57$\pm$1.45          & 70.77$\pm$1.17             \\
INCV             & 80.34$\pm$0.36          & 92.63$\pm$0.11          & 74.11$\pm$0.43               & 76.22$\pm$0.22          & 73.06$\pm$0.36             \\
OUSM        & 80.53$\pm$1.10          & 93.69$\pm$0.42          & 73.70$\pm$0.96               & 74.81$\pm$1.76          & 73.27$\pm$0.39             \\
NF-Net           & 51.23$\pm$1.18          & 69.92$\pm$1.17          & 33.21$\pm$0.39               & 37.19$\pm$0.94          & 36.27$\pm$0.62             \\
DM        & 77.25$\pm$0.21          & 87.58$\pm$0.36          & 69.78$\pm$0.32               & 70.68$\pm$0.23          & 69.11$\pm$0.38             \\
SELF             & 80.49$\pm$0.42          & 93.99$\pm$0.58          & 75.31$\pm$0.63               & 76.14$\pm$0.78          & 74.69$\pm$0.59             \\
\textbf{Ours}    & \textbf{83.40$\pm$0.20} & \textbf{94.51$\pm$0.34} & \textbf{76.54$\pm$0.33}      & \textbf{78.33$\pm$0.30} & \textbf{75.45$\pm$0.42}    \\ \bottomrule
\end{tabular}
\end{table}

\begin{figure}[]
  \centering
  \includegraphics[width=2.9in,height=8.667in]{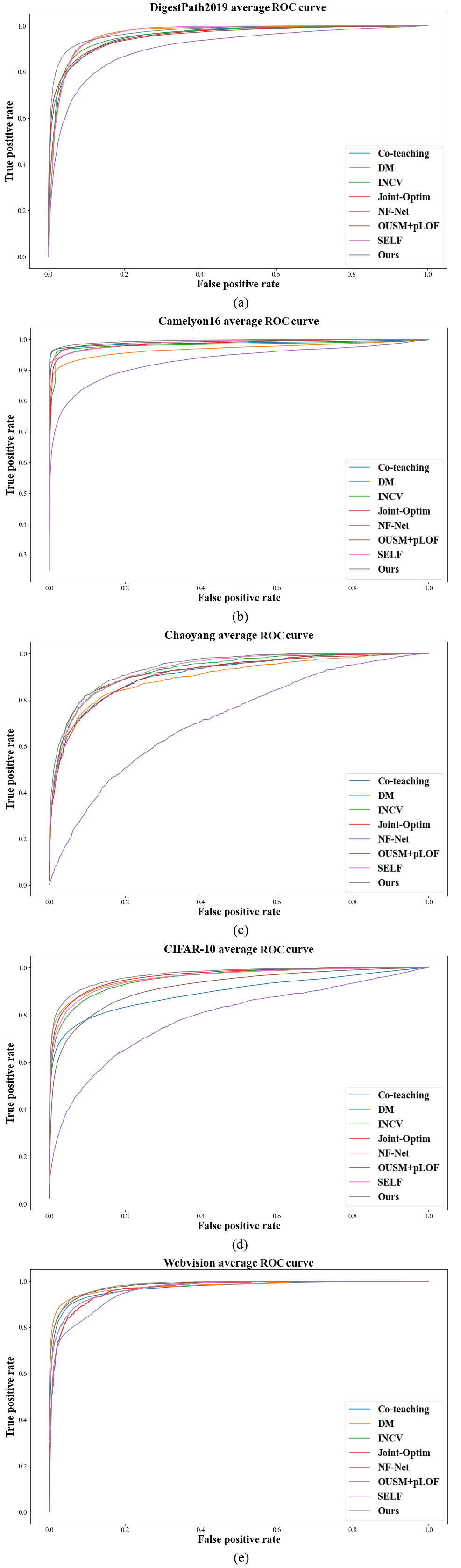}
  \caption{{(a) DigestPath2019 dataset average {ROC} curve from 10\% to 40\% noise ratios (used marco-average). (b) Camelyon16 dataset average {ROC} curve from 10\% to 40\% noise ratios (used marco-average). (c) Chaoyang dataset average {ROC} curve from 4 classes (used marco-average). (d) CIFAR-10 dataset average {ROC} curve from 10 classes with 20\% to 80\% noise ratios (used marco-average). (e) Webvision dataset average {ROC} curve from 50 classes (used marco-average).}}
  \label{fig:roc}
\end{figure}

\begin{table*}[]
\scriptsize
\centering
\caption{{Average test ACC, F1 Score, AUC, {Precision}, Recall(\%, 3 runs) with standard deviation on CIFAR-10 dataset.}}
\label{table:objCIFAR}
\begin{tabular}{@{}lcccccccccc@{}}
\toprule
\multicolumn{1}{c}{Noise ratio} & \multicolumn{5}{c}{20\% Sym.}                                                                                    & \multicolumn{5}{c}{50\% Sym.}                                                                                    \\ \midrule
Method                & ACC                  & AUC                  & F1 Score             & {Precision}                  & Recall               & ACC                  & AUC                  & F1 Score             & {Precision}                  & Recall               \\ \midrule
Joint                     & 88.37$\pm$0.06          & 98.88$\pm$0.06          & 88.33$\pm$0.03          & 88.38$\pm$0.03          & 88.37$\pm$0.06          & 81.85$\pm$0.16          & 97.88$\pm$0.06          & 81.87$\pm$0.16          & 82.03$\pm$0.12          & 81.85$\pm$0.16          \\
Co-tea.                     & 87.80$\pm$0.32           & 99.08$\pm$0.02          & 87.85$\pm$0.31          & 87.95$\pm$0.30          & 87.80$\pm$0.32           & 83.21$\pm$0.48          & 98.29$\pm$0.04          & 83.27$\pm$0.44          & 83.48$\pm$0.41          & 83.21$\pm$0.48          \\
INCV                            & 89.02$\pm$0.24          & 99.10$\pm$0.05           & 89.08$\pm$0.24          & 89.89$\pm$0.22          & 89.02$\pm$0.24          & 84.98$\pm$0.32          & 98.38$\pm$0.09          & 85.10$\pm$0.33           & 85.88$\pm$0.33          & 84.98$\pm$0.32          \\
OUSM                       & 81.88$\pm$0.20          & 97.01$\pm$0.06          & 81.85$\pm$0.24          & 81.88$\pm$0.25          & 81.88$\pm$0.20          & 68.66$\pm$0.81          & 93.77$\pm$0.23          & 68.50$\pm$0.77          & 69.39$\pm$0.52          & 68.66$\pm$0.81          \\
NF-Net                          & 80.27$\pm$1.15          & 96.76$\pm$0.50          & 80.19$\pm$1.03          & 81.02$\pm$1.21          & 80.27$\pm$1.15          & 67.38$\pm$1.24          & 90.34$\pm$0.34          & 67.58$\pm$1.11          & 67.88$\pm$1.07          & 67.38$\pm$1.24          \\
DM                       & 91.76$\pm$0.09          & 99.18$\pm$0.06          & 91.71$\pm$0.10          & 91.76$\pm$0.12          & 91.76$\pm$0.09          & 90.46$\pm$0.27          & 98.85$\pm$0.01          & 90.43$\pm$0.29          & 90.46$\pm$0.28          & 90.46$\pm$0.27          \\
SELF                            & 91.01$\pm$0.12          & 99.15$\pm$0.09          & 91.29$\pm$0.22          & 91.58$\pm$0.23          & 91.01$\pm$0.12          & 90.29$\pm$0.44          & 98.54$\pm$0.10          & 90.43$\pm$0.23          & 90.51$\pm$0.27          & 90.29$\pm$0.44          \\
\textbf{Ours}                   & \textbf{92.35$\pm$0.09} & \textbf{99.23$\pm$0.07} & \textbf{92.43$\pm$0.10} & \textbf{92.50$\pm$0.17} & \textbf{92.35$\pm$0.09} & \textbf{91.33$\pm$0.09} & \textbf{99.07$\pm$0.06} & \textbf{91.36$\pm$0.09} & \textbf{91.50$\pm$0.07} & \textbf{91.33$\pm$0.09} \\ \midrule
\multicolumn{1}{c}{Noise ratio} & \multicolumn{5}{c}{80\% Sym.}                                                                                    & \multicolumn{5}{c}{40\% Asym.}                                                                                   \\ \midrule
Method                & ACC                  & AUC                  & F1 Score             & {Precision}                  & Recall               & ACC                  & AUC                  & F1 Score             & {Precision}                  & Recall               \\ \midrule
Joint                     & 57.83$\pm$1.93          & 89.62$\pm$0.95          & 57.90$\pm$1.96          & 57.93$\pm$2.02          & 57.84$\pm$1.93          & 87.05$\pm$0.32          & 98.03$\pm$0.12          & 87.04$\pm$0.31          & 87.09$\pm$0.37          & 87.05$\pm$0.32          \\
Co-tea.                     & 24.37$\pm$4.21          & 61.97$\pm$2.88          & 24.92$\pm$4.52          & 25.68$\pm$4.02          & 24.37$\pm$4.21          & 82.86$\pm$1.13          & 98.15$\pm$0.28          & 82.93$\pm$1.01          & 83.24$\pm$0.79          & 82.86$\pm$1.13          \\
INCV                            & 53.98$\pm$2.68          & 85.76$\pm$1.23          & 53.98$\pm$2.35          & 53.99$\pm$2.24          & 53.98$\pm$2.68          & 85.88$\pm$0.67          & 98.25$\pm$0.09          & 85.9$\pm$0.54           & 85.99$\pm$0.52          & 85.88$\pm$0.67          \\
OUSM                       & 40.62$\pm$1.03          & 81.18$\pm$0.97          & 40.59$\pm$1.24          & 40.87$\pm$1.46          & 40.42$\pm$1.03          & 73.87$\pm$1.20          & 95.87$\pm$0.10          & 73.24$\pm$1.29          & 75.14$\pm$0.47          & 72.87$\pm$1.20          \\
NF-Net                          & 18.28$\pm$3.31          & 56.30$\pm$1.89          & 18.90$\pm$3.65          & 19.78$\pm$3.72          & 18.28$\pm$3.31          & 69.98$\pm$1.11          & 91.43$\pm$0.18          & 70.02$\pm$1.03          & 70.16$\pm$1.12          & 69.98$\pm$1.11          \\
DM                       & 58.04$\pm$1.67          & 88.62$\pm$1.01          & 58.77$\pm$1.35          & 59.71$\pm$1.22          & 58.04$\pm$1.67          & 86.50$\pm$0.19          & 97.51$\pm$0.02          & 86.24$\pm$0.19          & 86.73$\pm$0.05          & 86.50$\pm$0.19          \\
SELF                            & 59.75$\pm$2.13          & 86.40$\pm$1.34          & 59.24$\pm$1.99          & 60.02$\pm$1.87          & 59.75$\pm$2.13          & 87.14$\pm$0.54          & 98.12$\pm$0.11          & 87.27$\pm$0.64          & 87.35$\pm$0.48          & 87.14$\pm$0.54          \\
\textbf{Ours}                   & \textbf{61.69$\pm$1.22} & \textbf{89.91$\pm$0.77} & \textbf{61.6$\pm$1.34}  & \textbf{61.72$\pm$1.45} & \textbf{61.69$\pm$1.22} & \textbf{88.26$\pm$0.12} & \textbf{98.77$\pm$0.03} & \textbf{88.24$\pm$0.10} & \textbf{88.29$\pm$0.07} & \textbf{88.26$\pm$0.12} \\ \bottomrule
\end{tabular}
\end{table*}

\begin{table}[]
\scriptsize
\centering
\caption{{Average test ACC, F1 Score, AUC, {Precision}, Recall(\%, 3 runs) with standard deviation on Webvision dataset.}}
\label{table:objWebvision}
\begin{tabular}{@{}llllll@{}}
\toprule
Method & \multicolumn{1}{c}{ACC} & \multicolumn{1}{c}{AUC} & \multicolumn{1}{c}{F1 Score} & \multicolumn{1}{c}{{Precision}} & \multicolumn{1}{c}{Recall} \\ \midrule
Joint      & 60.28$\pm$0.92             & 96.70$\pm$0.07             & 60.40$\pm$1.03                  & 67.51$\pm$1.22             & 60.28$\pm$0.92                \\
Co-tea.      & 63.77$\pm$1.31             & 96.98$\pm$0.09             & 63.92$\pm$1.46                  & 64.02$\pm$1.51             & 63.77$\pm$1.31                \\
INCV             & 65.02$\pm$0.86             & 97.11$\pm$0.04             & 65.11$\pm$0.98                  & 65.05$\pm$1.02             & 65.02$\pm$0.86                \\
OUSM        & 70.86$\pm$0.50             & 98.16$\pm$0.05             & 70.95$\pm$0.41                  & 73.05$\pm$0.50             & 70.86$\pm$0.50                \\
NF-Net           & 58.03$\pm$1.27             & 95.86$\pm$0.08             & 58.77$\pm$1.34                  & 59.01$\pm$1.54             & 58.03$\pm$1.27                \\
DM        & 76.92$\pm$0.24             & 97.69$\pm$0.02             & 76.98$\pm$0.27                  & 77.08$\pm$0.43             & 76.92$\pm$0.24                \\
SELF             & 69.18$\pm$0.32             & 97.17$\pm$0.05             & 69.21$\pm$0.38                  & 69.80$\pm$0.63             & 69.18$\pm$0.32                \\
\textbf{Ours}    & \textbf{77.52$\pm$0.51}    & \textbf{98.32$\pm$0.04}    & \textbf{77.58$\pm$0.59}         & \textbf{78.11$\pm$0.87}    & \textbf{77.52$\pm$0.51}       \\ \bottomrule
\end{tabular}
\end{table}

\begin{enumerate}

{\item \textbf{SELF}\cite{2020SELF} (Duc Tam Nguyen \textit{et al.} 2020) first obtains the self-ensemble predictions of all training samples and then progressively removes samples whose ensemble predictions do not agree with their annotated labels \cite{2020Learning}.}

\item \textbf{DM}{\cite{2020DivideMix}} (Junnan Li \textit{et al.} 2020) lets its two-component and one-dimensional Gaussian mixture model be fitted to the training loss to obtain the confidence of an annotated label. By setting a confidence threshold, the training data is categorized into a labeled set and an unlabeled set. Subsequently, \textbf{MixMatch} \cite{david2019mixmatch} is employed to train a DNN for the transformed data.
\item \textbf{INCV}{\cite{chen2019understanding}} (Pengfei Chen \textit{et al.} 2019) randomly divides noisy training data and then employs cross-validation to classify true-labeled samples while removing large-loss samples at each training round. 
\item \textbf{Joint}{\cite{tanaka2018joint}} (Tanaka \textit{et al.} 2018)  jointly optimizes the sample labels and the network parameters.
\item \textbf{Co-tea.}{\cite{han2018co}} (Han \textit{et al.} 2018) maintains two networks. Each network selects samples of small training loss from the mini-batches and feeds them to the other network. 

\item \textbf{NF-Net}{\cite{cao2020breast}} (Zhantao Cao \textit{et al.} 2020) adopts a double-softmax classification module to prevent deep models from overfitting the noisy labels and a teacher-student module to strengthen the effect of clean labels. 
\item \textbf{OUSM}{\cite{cheng2019robust}} (Cheng Xue \textit{et al.} 2019) proposes online uncertainty sample mining and individual re-weighting methods to train their network.

\end{enumerate}


{Among them, work 1) to 5) are the state-of-the-art methods for general noisy data processing in recent years; work 6) and work 7) are the state-of-the-art methods proposed for medical data scenarios. We choose these schemes to contrast with to fully prove the superiority of our method. The testings of 2) to 6) are based on the open-source codes from the authors. We re-implemented and tested 1) and 7) based on the settings from the original papers.}



{For experiments on medical scenarios datasets, Table \ref{table:objMiccai} and Table \ref{table:objCamelyon} shows the test ACC, AUC, F1 Score, {Precision}, Recall on DigestPath2019 and Camelyon16 with different levels of label noise ranging from 10\% to 40\%. Our method almost outperforms the state-of-the-art methods across all noise ratios. Table \ref{table:objChaoyang} shows these metrics on the Chaoyang dataset. Our method outperforms all other methods by a large margin in every criterion. 

For experiments on natural computer vision datasets, Table \ref{table:objCIFAR} shows the test ACC, AUC, F1 Score, {Precision}, Recall on CIFAR-10 with different levels and different types of label noise ranging from 20\% to 80\%. Our method outperforms the state-of-the-art methods across all noise ratios. Table \ref{table:objWebvision} shows these metrics on Webvision dataset. Our method consistently outperforms all other methods. Besides, we show the mean {ROC} curves of all five datasets in Fig. \ref{fig:roc}.}


\begin{table}[]
\scriptsize
\centering
\caption{Average ACC(\%, 3 runs) with standard deviation of different ablation study on the DigestPath2019 dataset.}
\label{table:albMiccai}
\begin{tabular}{lllll}
\toprule
\multicolumn{1}{c}{Noise ratio}        & \multicolumn{1}{c}{10\%} & \multicolumn{1}{c}{20\%} & \multicolumn{1}{c}{30\%} & \multicolumn{1}{c}{40\%} \\ \midrule
Method                 & \multicolumn{4}{c}{ACC}                                                                                     \\ \midrule
Ours                                       & \textbf{94.91$\pm$0.52}      & \textbf{94.46$\pm$0.20}      & \textbf{91.72$\pm$0.69}      & \textbf{87.15$\pm$0.41}     \\ 
w/o NSHE                       & 93.46$\pm$0.12               & 91.48$\pm$0.43               & 90.86$\pm$0.24               & 86.85$\pm$0.60              \\ 
w/o (NSHE + EHN)       & 90.54$\pm$0.45               & 88.32$\pm$0.17               & 86.25$\pm$0.48               & 85.15$\pm$1.87              \\ 
w/o whole & 90.01$\pm$0.49               & 87.24$\pm$0.98               & 85.11$\pm$1.31               & 82.40$\pm$1.42              \\ \bottomrule
\end{tabular}
\end{table}

\begin{table}[]
\scriptsize
\centering
\caption{Average ACC(\%, 3 runs) with standard deviation of different ablation study on the Camelyon16 dataset.}
\label{table:albCame}
\begin{tabular}{lllll}
\toprule
\multicolumn{1}{c}{Noise ratio}            & \multicolumn{1}{c}{10\%} & \multicolumn{1}{c}{20\%} & \multicolumn{1}{c}{30\%} & \multicolumn{1}{c}{40\%} \\ \midrule
Method                 & \multicolumn{4}{c}{ACC}                                                                                     \\ \midrule
Ours                                       & \textbf{98.82$\pm$0.20}      & \textbf{98.61$\pm$0.11}      & \textbf{98.32$\pm$0.41}      & \textbf{98.17$\pm$0.24}     \\ 
w/o NSHE                        & 97.15$\pm$0.50               & 97.00$\pm$0.98               & 96.83$\pm$0.89               & 96.55$\pm$2.11              \\ 
w/o (NSHE + EHN)       & 95.99$\pm$0.73               & 94.95$\pm$1.13               & 93.78$\pm$1.54               & 93.45$\pm$2.44              \\ 
w/o whole & 94.60$\pm$0.66               & 93.34$\pm$0.91               & 93.11$\pm$0.78               & 92.23$\pm$1.82              \\ \bottomrule
\end{tabular}
\end{table}

\subsection{Ablation Study}

\begin{figure}[]
  \centering
  \includegraphics[width=3.48in]{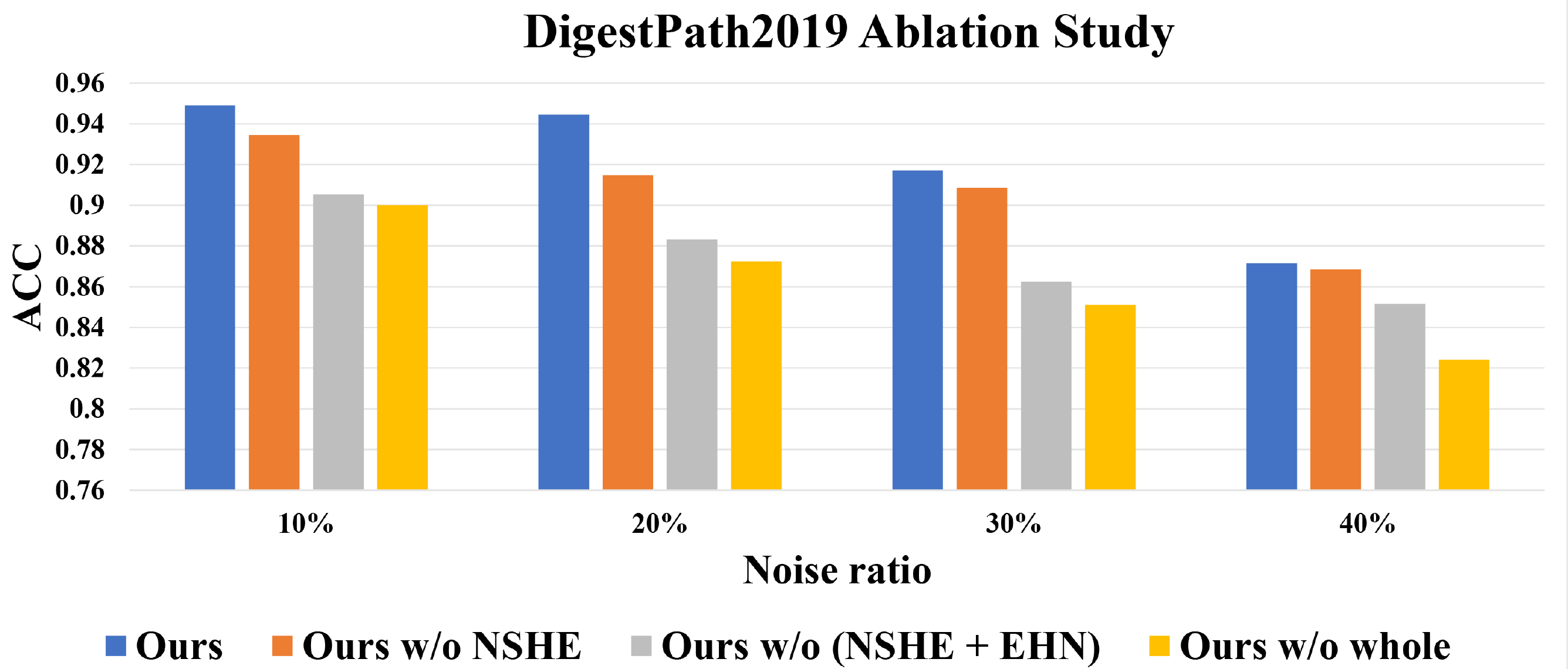}
  \caption{{Ablation study results in terms of test accuracy on DigestPath2019.}}
  \label{fig:albMiccai}
\end{figure}

\begin{figure}[]
  \centering
  \includegraphics[width=3.48in]{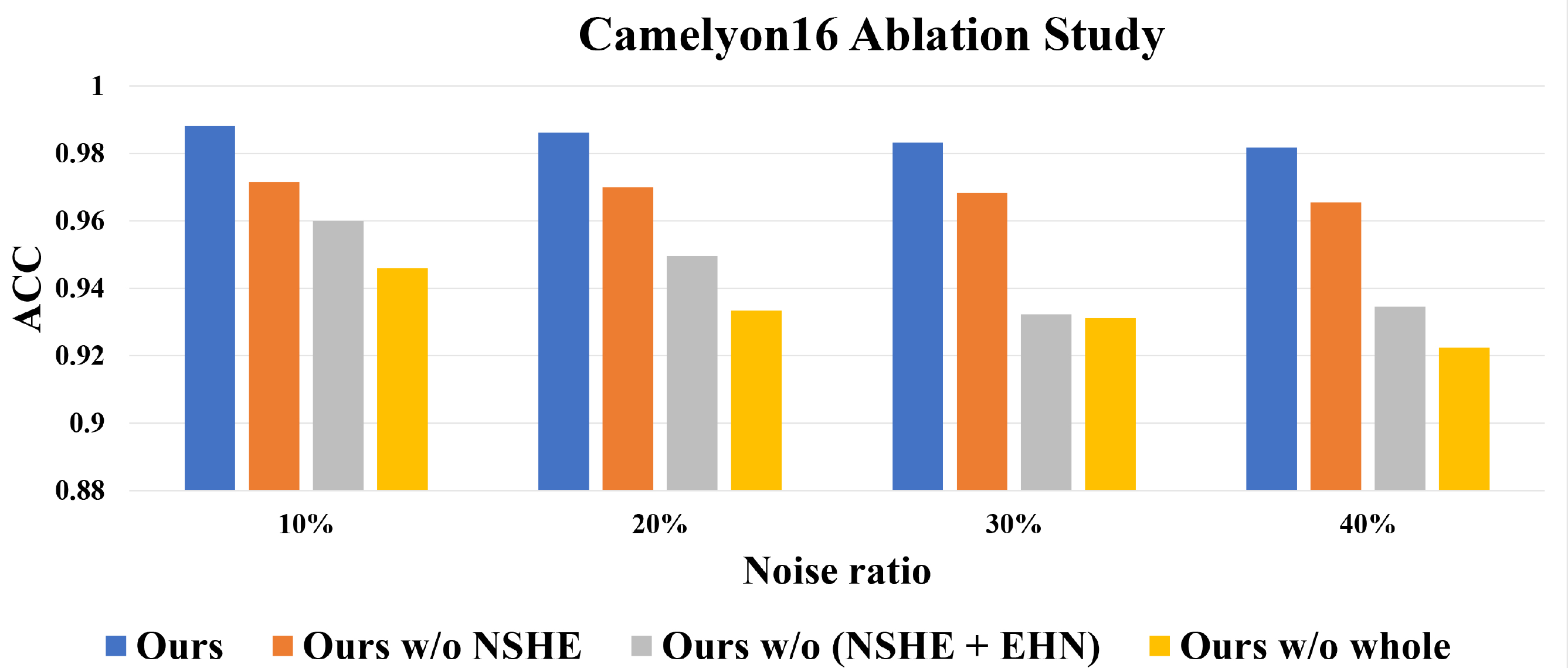}
  \caption{{Ablation study results in terms of test accuracy on Camelyon16 dataset.}}
  \label{fig:albCame}
\end{figure}

We study the effect of removing different components to provide insights into what makes our method successful. Fig. \ref{fig:albMiccai} and Fig. \ref{fig:albCame} show the ablation study results in different noise ratios. The result details are shown in Table \ref{table:albMiccai} to \ref{table:albCame}, and we discuss them below.


To study the effects of the NSHE scheme, we removed the NSHE scheme (w/o NSHE), namely train the single model by using the dataset from label correction. The results show the hard samples play quite a significant role in training final models. By removing the NSHE scheme, the test accuracy decreased by an average of about 1.5\%.

{To study the effects of the EHN detection scheme, we removed both the EHN detection scheme and NSHE scheme (w/o (EHN + NSHE)). In this situation, following work \cite{tanaka2018joint}, we directly used the classification model as the correction model, and the dataset is processed only by the correction model. The results show the EHN detection scheme is very effective to save more hard samples and filtered more noisy ones. Without the EHN detection scheme, the test accuracy further decreased by an average of about 2.8\%.}

{To study the effects of the correction model, we further removed the whole label correction phase (w/o whole), namely we train the model by using original data. To converge the model under the same epoch, we adjusted the learning rate, which would smooth the test accuracy in the last epoch but the highest accuracy will be affected. By removing the correction model, the test accuracy further decreased by an average of about 1.2\%.}

Among the NSHE scheme, EHN detection scheme, and correction model, the EHN detection scheme introduces the maximum performance gain, followed by the NSHE scheme. All components have a certain gain at any noise ratio.

\begin{figure}[]
  \centering
  \includegraphics[width=2.7in]{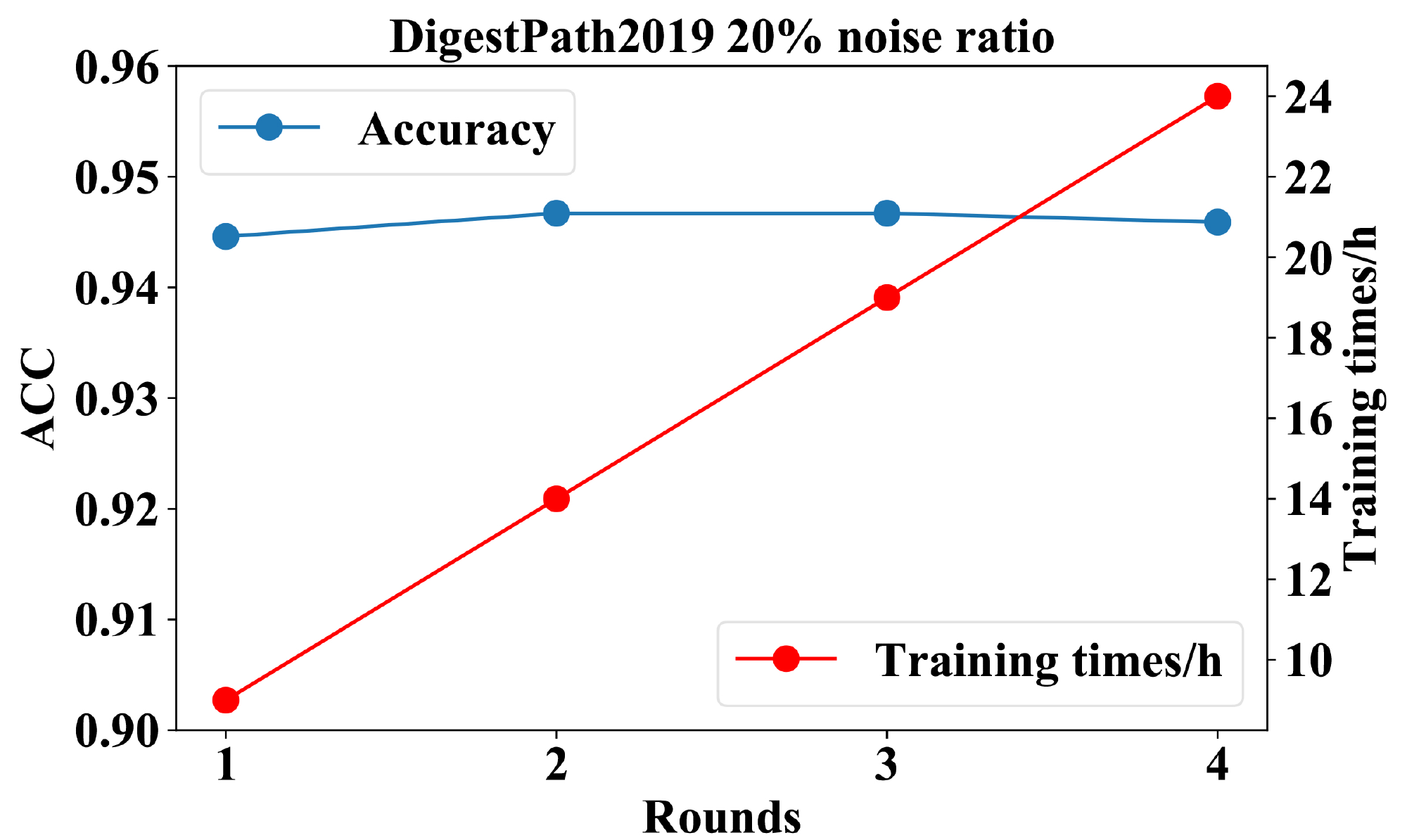}
  \caption{{Our method with different rounds (just first stage) results in terms of test accuracy on DigestPath2019 dataset with 20\% noise ratio.}}
  \label{fig:alb2}
\end{figure}

\begin{figure}[]
  \centering
  \includegraphics[width=2.7in]{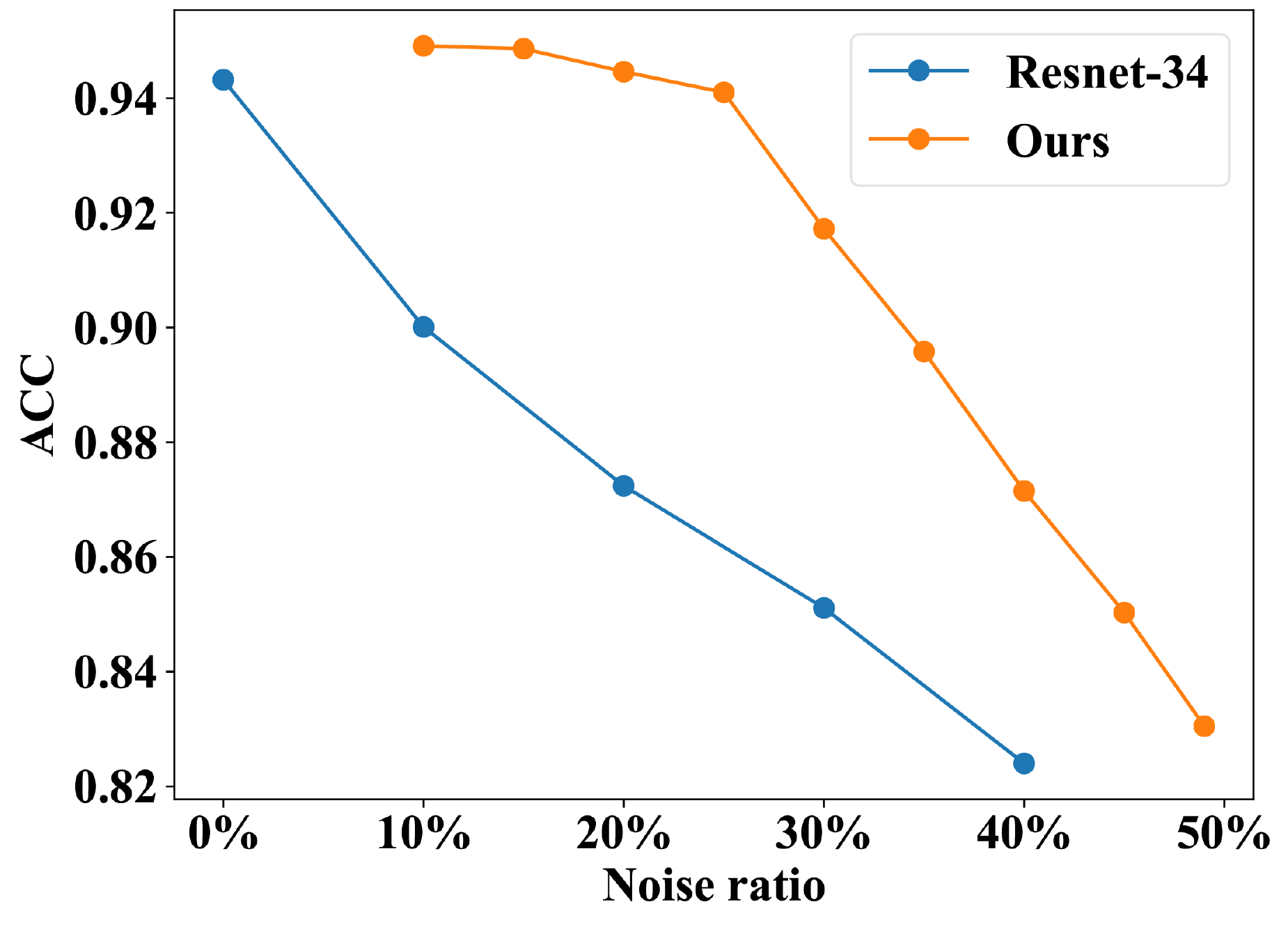}
  \caption{{Test accuracy of a simple Resnet-34 and our algorithm (backbone is Resnet-34) on DigestPath2019 dataset with different noise ratios.}}
  \label{fig:simple_ours}
\end{figure}

{To analyze the effect of self-training rounds, we recorded the training times and test accuracy in different self-training rounds. Fig. \ref{fig:alb2} shows the results on DigestPath2019 dataset with 20\% noise ratio. Training more rounds consumes more computing resources but brings little gain. Therefore, we choose to train only one round in our experiment. We also studied how much higher noise ratio could our scheme tolerate to reach a similar performance against a simple model. We first directly trained a simple Resnet-34 under 5 different noise ratios (0\%, 10\%, 20\%, 30\%, 40\%, respectively) on DigestPath2019 dataset. Then we also carried out experiments }{our algorithm (backbone is Resnet-34) at 10\%, 15\%, 20\%, 25\%, 30\%, 35\%, 40\%, 45\%, 49\% noise ratios (note that we select 49\% noise ratio because it is close to the theoretical limit of 50\% for this type of noise on binary classification task). The results are shown in Fig. \ref{fig:simple_ours}.}

{According to Fig. \ref{fig:simple_ours}, to reach the same performance, for 0\%, 10\%, 20\%, 30\%, and 40\% noise ratios, our scheme can tolerate about up to 22\%, 34\%, 40\%, 44\%, and 49\% noise ratio, respectively. We also find that in the noise ratios of 0\% to 20\% on DigestPath2019 dataset, the results of our method are even better than the trained simple Resnet-34 on a completely clean dataset. We believe that this phenomenon is due to our enhancement of the information of hard samples.}

\subsection{Analysis of EHN Detection Scheme}

\begin{figure}[]
  \centering
  \includegraphics[width=2.6in]{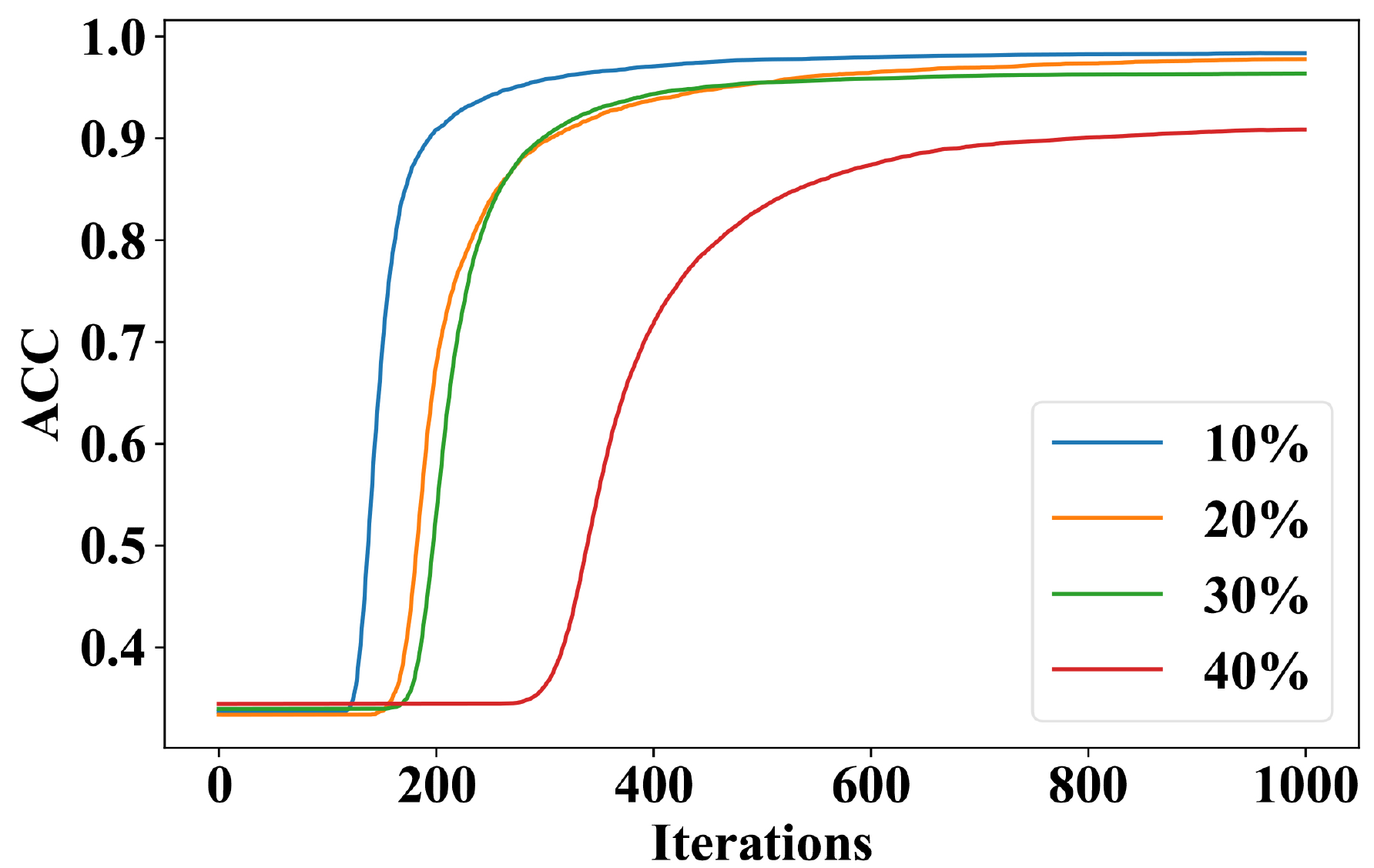}
  \caption{{Accuracy vs. iterations graph of $M_m$ in DigestPath2019 dataset with different noise ratios.}}
  \label{fig:ehn_acc}
\end{figure}

\begin{figure}[]
  \centering
  \includegraphics[width=3.55in]{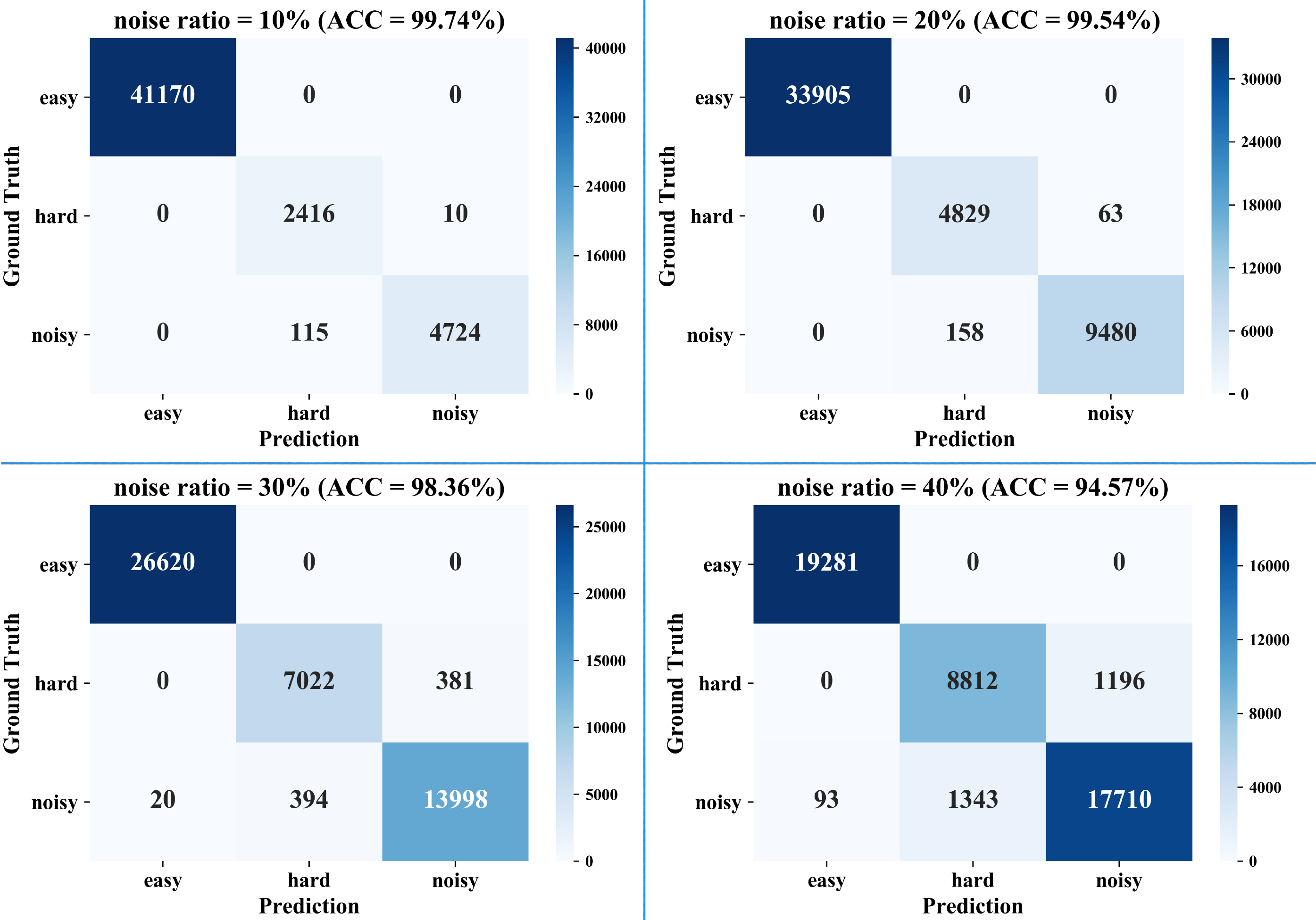}
  \caption{{Confusion matrix and accuracy (ACC) of EHN detection scheme in DigestPath2019 dataset with different noise ratios.}}
  \label{fig:EHN_diffNoiseR}
\end{figure}

{\textbf{Effectiveness and Convergence Analysis.} To analyze the effectiveness and convergence of our EHN detection scheme, we trained on DigestPath2019 dataset with different noise ratios and plotted the curves of test ACC vs. Iterations (``test'' set here means $D \setminus D_e$) of $M_m$. The results are shown in Fig. \ref{fig:ehn_acc}. According to Fig. \ref{fig:ehn_acc}, $M_m$ has converged in the later training stage at each noise ratio, so the convergence of $M_m$ is relatively stable. Also from Fig. \ref{fig:ehn_acc}, $M_m$ can achieve 98\%, 97\%, 96\%, and 90\% classification accuracy in 10\% to 40\% noise ratios, respectively. Thus, $M_m$ can effectively distinguish the hard and noisy samples. We also recorded the confusion matrix of EHN detection scheme as Fig. \ref{fig:EHN_diffNoiseR} {(clean samples} {are divided into easy and hard in this test, as depicted by Fig. \ref{fig:meanHistorgam} (b)).}}

\begin{figure*}[]
  \centering
  \includegraphics[width=5.8in]{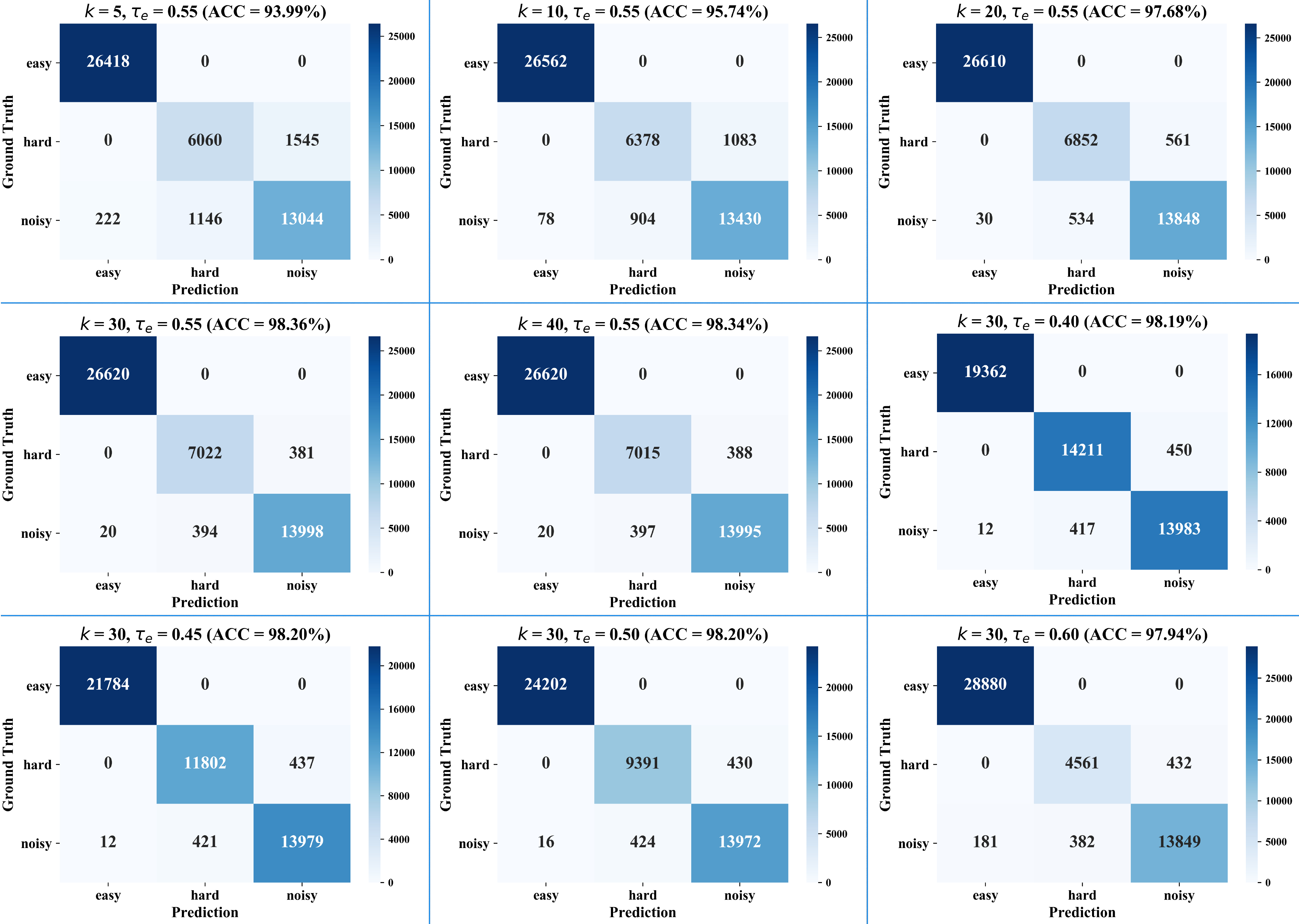}
  \caption{{Confusion matrix and accuracy (ACC) of EHN detection scheme on different ``{$k$}'' and ``$\tau_e$'' in DigestPath2019 dataset with 30\% noise ratio.}}
  \label{fig:EHN_diffPar}
\end{figure*}

\begin{figure}[]
  \centering
  \includegraphics[width=3.5in]{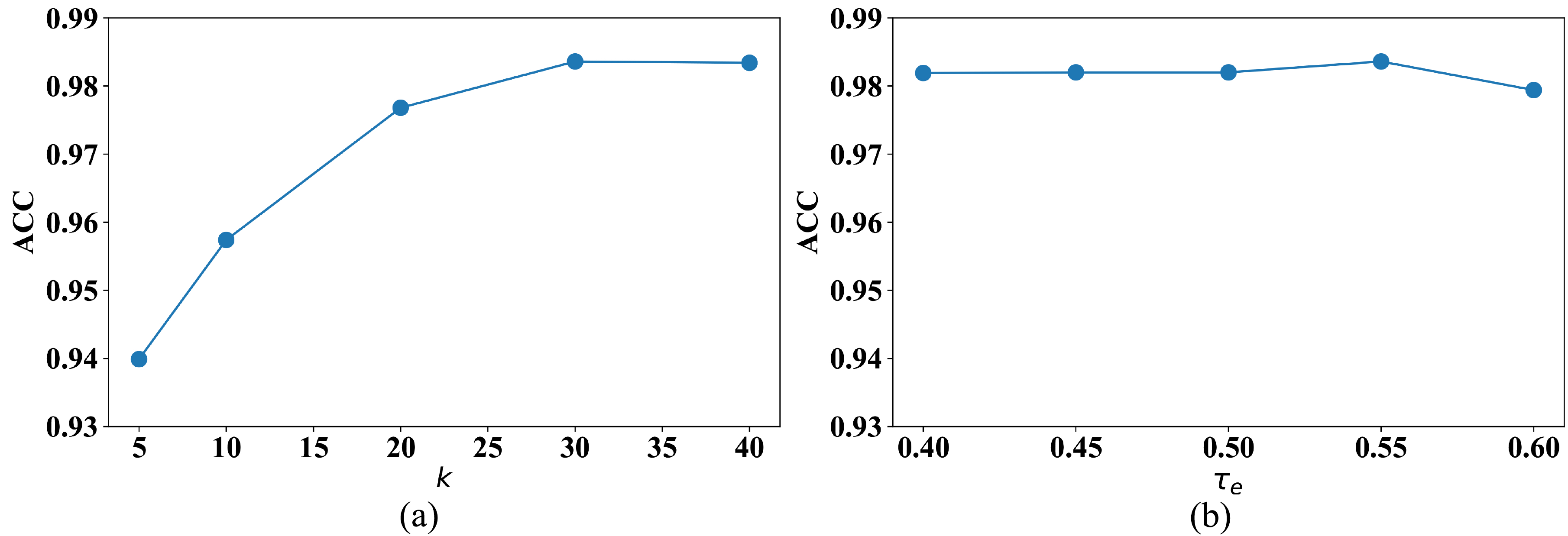}
  \caption{{(a) ACC vs. {$k$} graph of EHN detection scheme. (b) ACC vs. $\tau_e$ graph of EHN detection scheme.}}
  \label{fig:EHN_n}
\end{figure}

\begin{figure}[]
  \centering
  \includegraphics[width=3.5in]{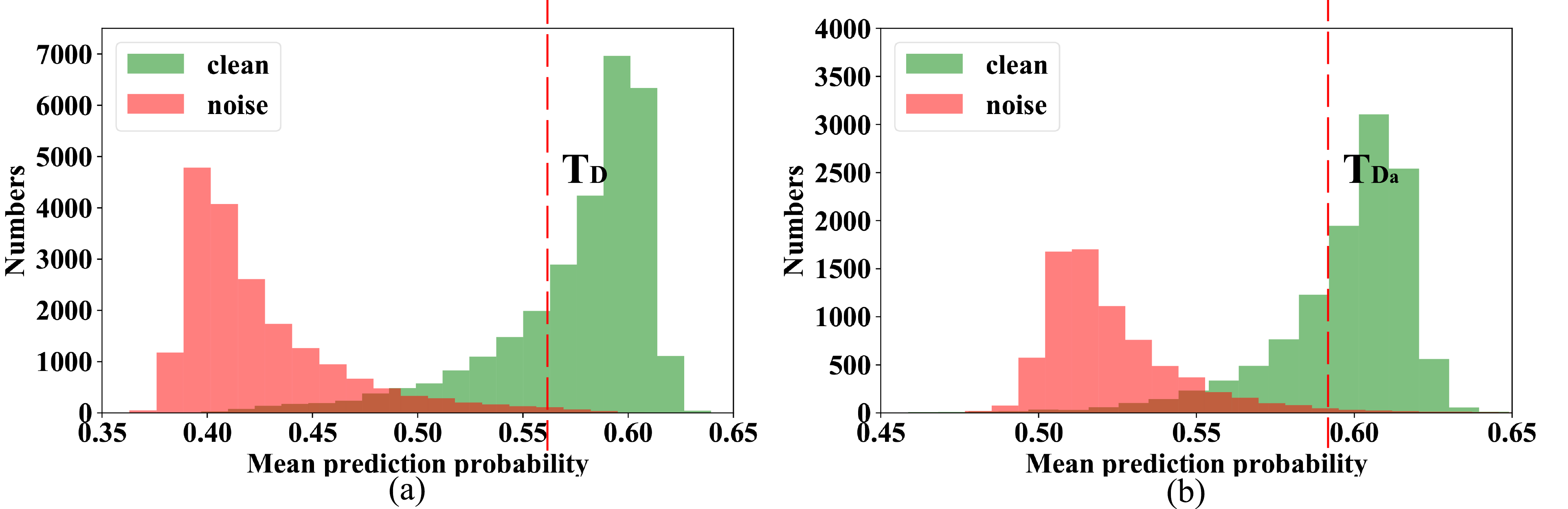}
  \caption{{(a) Mean prediction probability histogram of the clean and noisy samples in DigestPath2019 dataset (40\% noise ratio) $D$. (b) Mean prediction probability histogram of the clean and noisy samples in corresponding $D_a$. }}
  \label{fig:clean_noise}
\end{figure}

\textbf{{Parameter Analysis.}} {To analyze how sensitive the EHN detection scheme is to the training epochs {$k$} and easy sample ratio $\tau_e$, we trained on different {$k$} and $\tau_e$ in DigestPath2019 dataset with 30\% noise ratio, recorded the confusion matrix, and calculated the ACC of EHN detection scheme. Specifically, we first adjusted the value of {$k$} with fixed $\tau_e$ = 0.55, and thus obtained the sensitivity of EHN to {$k$}. Then we adjusted the value of $\tau_e$ with fixed {$k$} = 30, and thus obtained the sensitivity of EHN to $\tau_e$. There are a total of 9 different parameter configurations. The results are shown in Fig. \ref{fig:EHN_diffPar}. And we also plotted the ACC vs. Parameters graph as Fig. \ref{fig:EHN_n}. These results show that as {$k$} increases from 0, the performance continues to increase. When {$k$} reaches about 30, the performance achieves maximum accuracy. However, as {$k$} increases further, the performance begins to decrease slightly. But overall, when {$k$} exceeds a certain threshold, the performance of EHN is relatively stable. And for the parameter $\tau_e$, changing the value within a reasonable range has little impact on the performance. By the way, our utilized parameters are effective according to Fig. \ref{fig:EHN_n}.}

{\textbf{Why $D_a$ works?} To study why $M_m$ trained on the artificial created $D_a$ can recognize hard and noisy samples in original dataset $D$, we plotted both the mean prediction probability histogram of the clean and noisy samples in DigestPath2019 dataset (40\% noise ratio) $D$ and the corresponding $D_a$. The results are shown in Fig. \ref{fig:clean_noise}. According to Fig. \ref{fig:clean_noise} (b), there are also some clean samples which can not be distinguished from the noisy ones by mean prediction probability. Although the samples in $D_e$ are all easy ones to dataset $D$, part of the samples became hard ones to dataset $D_a$. In Fig. \ref{fig:clean_noise}, it should be noted that the mean prediction probabilities of samples trained on $D_a$ have similar distribution with the original dataset $D$, and this is why our EHN detection scheme works by using the artificial created $D_a$.}

\subsection{Hard and Noisy Sample Analysis}

\begin{figure}[]
  \centering
  \includegraphics[width=2.8in]{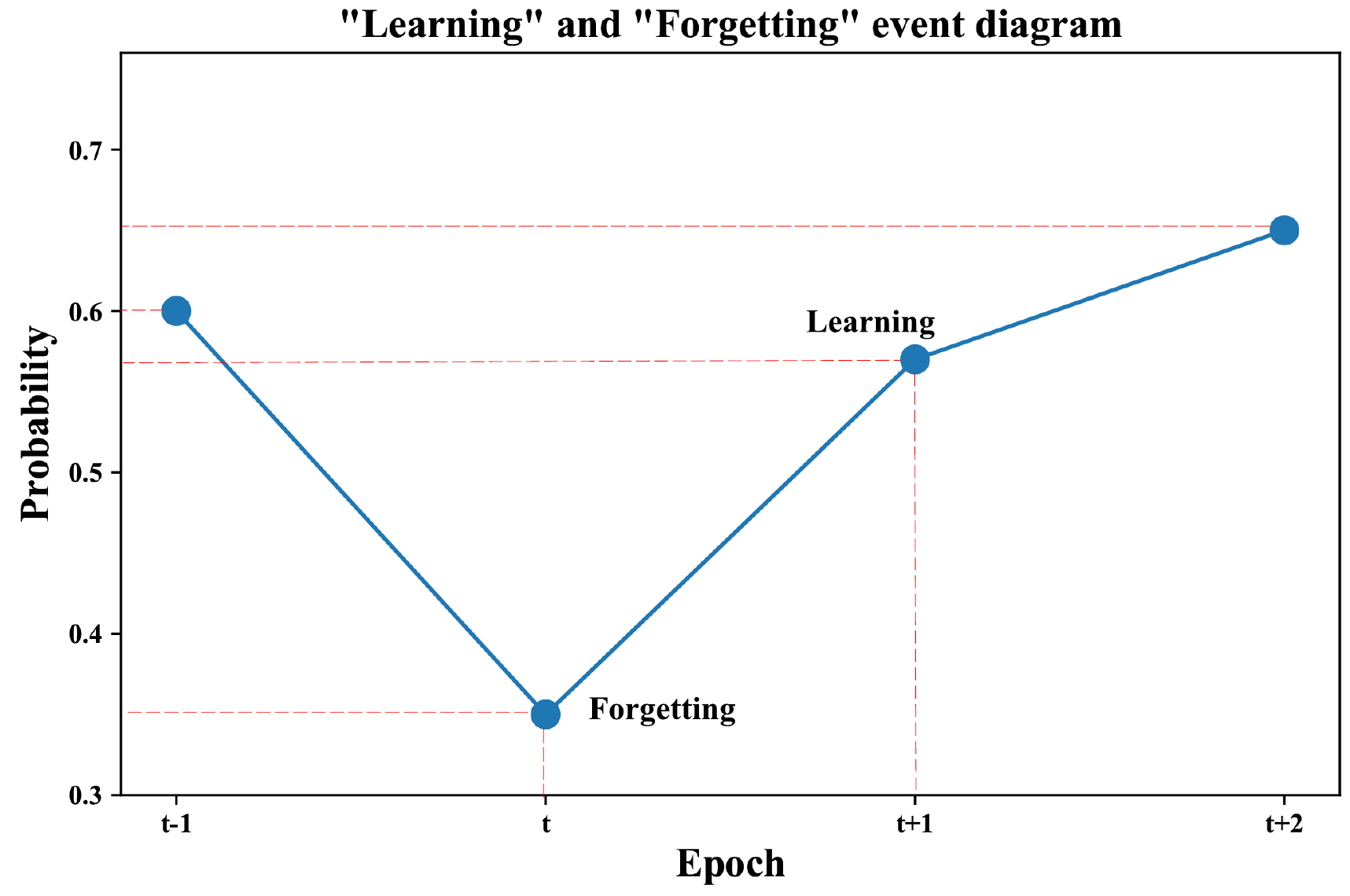}
  \caption{{The diagram of ``Learning'' and ``Forgetting'' event.}}
  \label{fig:fl_diagram}
\end{figure}

\begin{figure*}[hbt]
  \centering
  \includegraphics[width=6.6in]{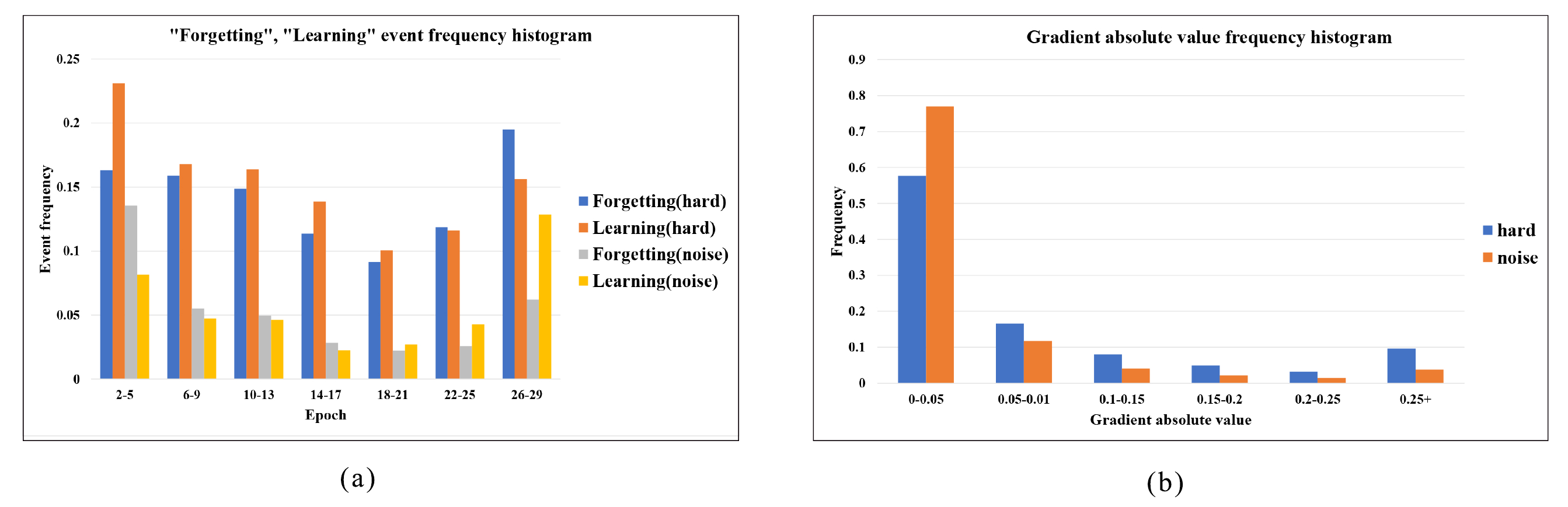}
  \caption{(a) The forgetting and learning event frequency histogram. (b) The gradient absolute value frequency histogram.} 
  \label{fig:hsa}
\end{figure*}

\textbf{Behavior Analysis.} To analyze the training process behavior of the hard sample and the noisy sample, inspired by \cite{toneva2018empirical}, we calculated the frequency of the ``Learning'' event and ``Forgetting'' event of them through the whole training epochs. The ``Learning'' event in the $t$ epoch is defined as an event that the prediction probability of the labeled class is less than $0.5$ in $t-1$ epoch, while greater than $0.5$ in $t$ epoch. The ``Forgetting'' event in the $t$ epoch is defined as an event that the prediction probability of the labeled class is greater than $0.5$ in $t-1$ epoch, while less than $0.5$ in $t$ epoch. Fig. \ref{fig:fl_diagram} shows the diagram of ``Learning'' and ``Forgetting'' event. The statistical results are shown in Fig. \ref{fig:hsa} (a). This results show that the hard sample and the noisy sample have great behavioral differences with the increase of the epoch in training. In the early stage of training, the hard samples tend to have more learning events, while the noisy samples tend to have more forgetting events. In the late stage of training, the hard samples tend to have more forgetting events while the noisy samples tend to have more learning events. On the whole, the frequency of learning and forgetting events of the hard samples is higher than that of the noisy samples during the whole training epochs.

\begin{figure}[]
  \centering
  \includegraphics[width=2.7in]{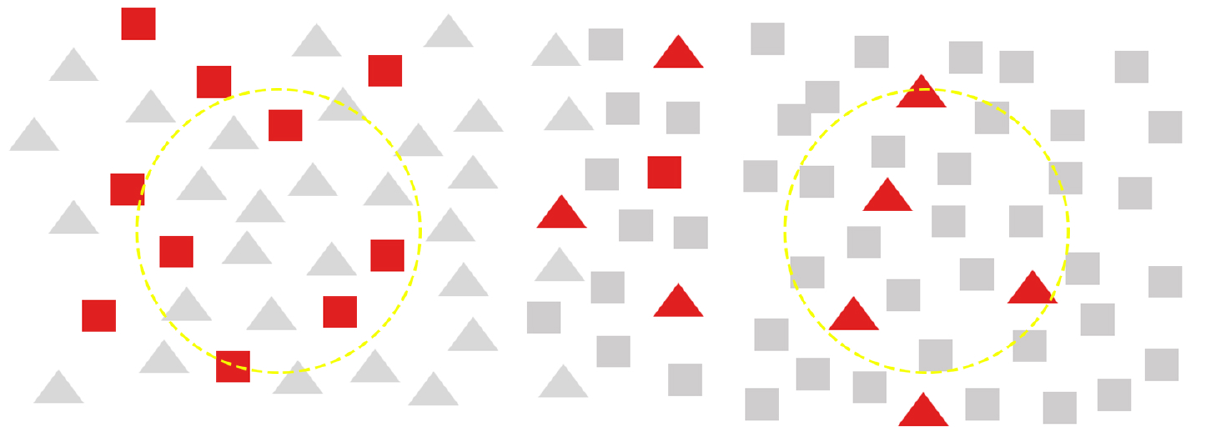}
  \caption{Sample distribution diagram, the red color represents the noisy sample. Yellow circles denote the center of clustering.}
  \label{fig:hsa_c}
\end{figure}

\begin{figure}[]
  \centering
  \includegraphics[width=3.5in]{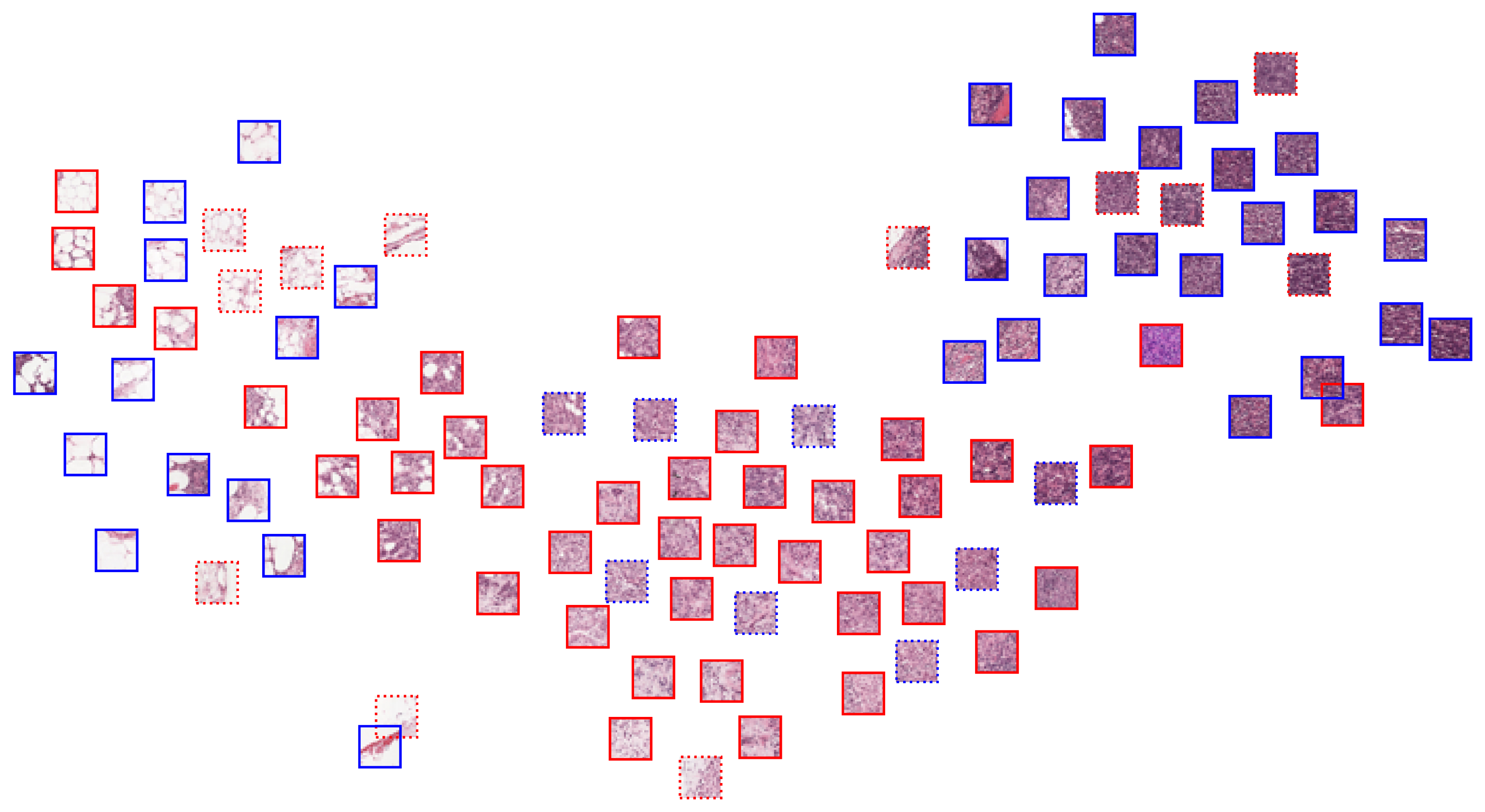}
  \caption{Visualization of Camelyon16 dataset with 20\% noise ratio by using t-SNE\cite{laurens2008visiualizing}, where patches with red solid border are clean malignant samples, patches with blue solid border are clean benign samples, patches with red dotted border are noisy malignant samples (true labels are benign), patches with blue dotted border are noisy benign samples (true labels are malignant).}
  \label{fig:tsne}
\end{figure}

\begin{figure}[hbt]
  \centering
  \includegraphics[width=3.5in]{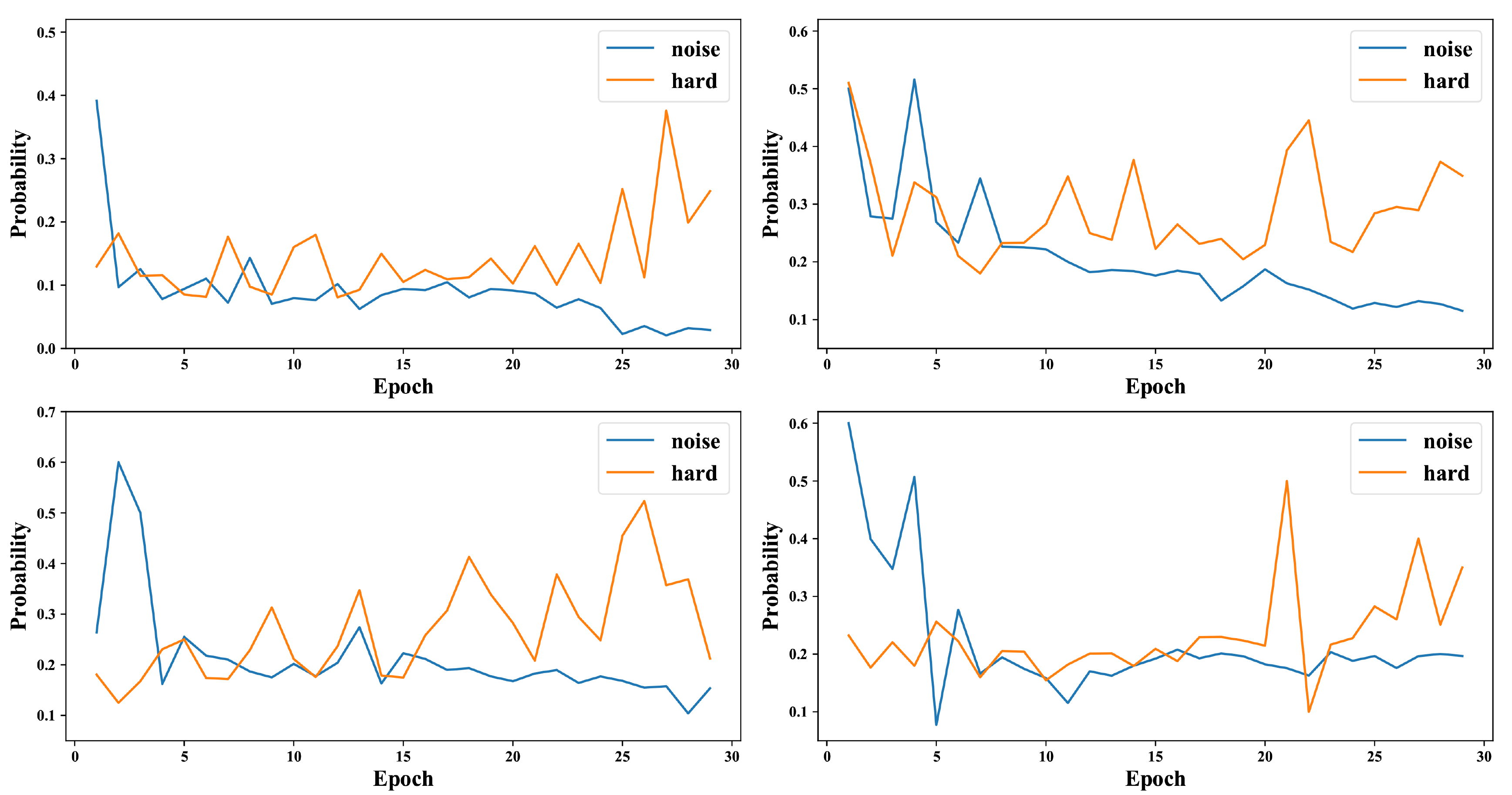}
  \caption{Training history graph of the selected hard and noisy samples.}
  \label{fig:hsa_d}
\end{figure}

\begin{table*}[]
\scriptsize
\centering
\caption{{The final noise ratios of different label correction schemes on DigestPath2019 dataset.}}
\label{table:hsa}
\begin{tabular}{@{}lcccccc@{}}
\toprule
Noise ratio  & \multicolumn{3}{c}{10\%}                                          & \multicolumn{3}{c}{20\%}                                          \\ \midrule
Strategy     & Remain samples         & Remain noisy samples & Final noise ratio & Remain samples         & Remain noisy samples & Final noise ratio \\ \midrule
Ours         & \multirow{2}{*}{46181} & 26                   & 0.0563\%          & \multirow{2}{*}{45895} & 159                  & 0.346\%           \\
Drop by mean &                        & 181                  & 0.392\%           &                        & 859                  & 1.87\%            \\ \midrule
Noise ratio  & \multicolumn{3}{c}{30\%}                                          & \multicolumn{3}{c}{40\%}                                          \\ \midrule
Strategy     & Remain samples         & Remain noisy samples & Final noise ratio & Remain samples         & Remain noisy samples & Final noise ratio \\ \midrule
Ours         & \multirow{2}{*}{42361} & 551                  & 1.30\%            & \multirow{2}{*}{39348} & 2227                 & 5.66\%            \\
Drop by mean &                        & 1537                 & 3.63\%            &                        & 6626                 & 16.8\%            \\ \bottomrule
\end{tabular}
\end{table*}

We also calculated the frequency histogram of the gradient absolute value. The gradient absolute value is the absolute value of the gradient between adjacent epochs, that is, the absolute value of the difference between the prediction probabilities of adjacent epochs. As shown in Fig. \ref{fig:hsa} (b), the gradient absolute value of the hard samples tends to be higher than the noisy samples. We believe that the reason for this phenomenon is that the hard samples' hard fitting attribute makes the prediction probability of the model jump frequently. Some noisy samples, however, are conspicuously at the center of other classes; they are ``super hard" for the model to fit. Their effect on the optimization of model parameters would be suppressed by the clean samples around. So in the latter part of the training process, their output probabilities would not change much. As shown in Fig. \ref{fig:hsa_c}, the noisy samples are scattered throughout the dataset. When these samples fall in the intersection area of two categories, their training behaviors are more similar to the hard ones; when these samples fall in the center area of other classes (yellow circles in Fig. \ref{fig:hsa_c}), they have distinct training behavior differences with the hard samples. To further prove the conjecture above, we selected samples in the Camelyon16 dataset with 20\% noise ratio and used the Resnet-34 model (pre-trained in ImageNet) to extract the features and visualized them by t-SNE\cite{laurens2008visiualizing} in Fig. \ref{fig:tsne}. We selected the noisy samples that fall in the center of other classes in Fig. \ref{fig:tsne} and plotted their training history to compare with the hard samples in Fig. \ref{fig:hsa_d}. It can be seen that these noisy samples are indeed difficult to predict in the latter part of the training, which confirms our previous analysis.


\textbf{Benefit Analysis.} To show the gain of our hard-aware label correction phase more intuitively, we count the remaining noise of the processed dataset and compare it with the baseline in Table \ref{table:hsa}. The baseline is set to update the labels by classification model and drop out the samples by using the mean prediction probability of the training history. The results show that with the same number of remaining samples, our strategy eliminates more noisy samples by protecting hard samples as much as possible, and thus generates a higher quality data set for the second phase of training. {Besides, for the real medical scenario dataset ``Chaoyang'', our hard-aware label correction phase only drops out 191 samples (total 4021 samples), and this shows that our method reduces noise with little damage to the original dataset.}

\section{Conclusion}
\label{sec:conclusion}
Deep learning-based histopathology image classification can improve the diagnosing accuracy of cancer. It is difficult to collect a large clean dataset for training such a classification model. The existing noisy label correction methods fail to distinguish the hard samples from the noisy samples and thus ruin the model performance. In this study, we proposed a hard sample aware noise robust learning for histopathology image classification to save more clean samples, and thus boosted the model performance. We found that the training prediction history can be used to distinguish the hard samples and noisy samples. By integrating our EHN detection scheme into the noise removing, more hard clean samples can be saved. Besides, in our NSHE scheme under co-learning architecture, we adopted different parameter updating speed for the two models. This can further suppress the interference of noisy samples. Our results provide compelling performance for the noisy dataset, and the proposed method can be directly applied to the real-word noisy scenario. In the future, we will conduct hard sample aware semantic segmentation, such as malignant tissue segmentation for histopathology images. 

\bibliographystyle{IEEEtran}
\bibliography{IEEEabrv,IEEEexample}

\begin{thebibliography}{10}
\providecommand{\url}[1]{#1}
\csname url@samestyle\endcsname
\providecommand{\newblock}{\relax}
\providecommand{\bibinfo}[2]{#2}
\providecommand{\BIBentrySTDinterwordspacing}{\spaceskip=0pt\relax}
\providecommand{\BIBentryALTinterwordstretchfactor}{4}
\providecommand{\BIBentryALTinterwordspacing}{\spaceskip=\fontdimen2\font plus
\BIBentryALTinterwordstretchfactor\fontdimen3\font minus
  \fontdimen4\font\relax}
\providecommand{\BIBforeignlanguage}[2]{{%
\expandafter\ifx\csname l@#1\endcsname\relax
\typeout{** WARNING: IEEEtran.bst: No hyphenation pattern has been}%
\typeout{** loaded for the language `#1'. Using the pattern for}%
\typeout{** the default language instead.}%
\else
\language=\csname l@#1\endcsname
\fi
#2}}
\providecommand{\BIBdecl}{\relax}
\BIBdecl

\bibitem{schiffman2015early}
J.~D. Schiffman, P.~G. Fisher, and P.~Gibbs, ``Early detection of cancer: past,
  present, and future,'' \emph{American Society of Clinical Oncology
  Educational Book}, vol.~35, no.~1, pp. 57--65, 2015.

\bibitem{wang2019breast}
Z.~Wang, M.~Li, H.~Wang, H.~Jiang, Y.~Yao, H.~Zhang, and J.~Xin, ``Breast
  cancer detection using extreme learning machine based on feature fusion with
  cnn deep features,'' \emph{IEEE Access}, vol.~7, pp. 105\,146--105\,158,
  2019.

\bibitem{campanella2019clinical-grade}
G.~Campanella, M.~G. Hanna, L.~Geneslaw, A.~Miraflor, V.~Werneck Krauss~Silva,
  K.~J. Busam, E.~Brogi, V.~E. Reuter, D.~S. Klimstra, and T.~J. Fuchs,
  ``Clinical-grade computational pathology using weakly supervised deep
  learning on whole slide images,'' in \emph{Nature Medicine}, 2019, pp.
  1301--1309.

\bibitem{chen2019an}
P.~H.~C. Chen, K.~Gadepalli, R.~MacDonald, Y.~Liu, S.~Kadowaki, K.~Nagpal,
  T.~Kohlberger, J.~Dean, G.~S. Corrado, J.~D. Hipp, C.~H. Mermel, and M.~C.
  Stumpe, ``An augmented reality microscope with real-time artificial
  intelligence integration for cancer diagnosis,'' in \emph{Nature Medicine},
  2019, pp. 1453--1457.

\bibitem{coudray2018classification}
N.~Coudray, P.~S. Ocampo, T.~Sakellaropoulos, N.~Narula, M.~Snuderl, D.~Fenyö,
  A.~L. Moreira, N.~Razavian, and A.~Tsirigos, ``Classification and mutation
  prediction from non–small cell lung cancer histopathology images using deep
  learning,'' in \emph{Nature Medicine}, 2018, pp. 1559--1567.

\bibitem{tang2019interpretable}
Z.~Tang, K.~V. Chuang, C.~DeCarli, L.~W. Jin, L.~Beckett, M.~J. Keiser, and
  B.~N. Dugger, ``Interpretable classification of alzheimer’s disease
  pathologies with a convolutional neural network pipeline,'' in \emph{Nature
  Communications}, 2018, pp. 1--14.

\bibitem{yi2019probabilistic}
K.~Yi and J.~Wu, ``Probabilistic end-to-end noise correction for learning with
  noisy labels,'' \emph{arXiv preprint arXiv:1903.07788}, 2019.

\bibitem{wei2019harnessing}
Y.~Wei, C.~Gong, S.~Chen, T.~Liu, J.~Yang, and D.~Tao, ``Harnessing side
  information for classification under label noise,'' \emph{IEEE Transactions
  on Neural Networks and Learning Systems}, 2019.

\bibitem{goldberger2016training}
J.~Goldberger and E.~Ben-Reuven, ``Training deep neural-networks using a noise
  adaptation layer,'' 2016.

\bibitem{hendrycks2018using}
D.~Hendrycks, M.~Mazeika, D.~Wilson, and K.~Gimpel, ``Using trusted data to
  train deep networks on labels corrupted by severe noise,'' in \emph{Advances
  in neural information processing systems}, 2018, pp. 10\,456--10\,465.

\bibitem{patrini2017making}
G.~Patrini, A.~Rozza, A.~Krishna~Menon, R.~Nock, and L.~Qu, ``Making deep
  neural networks robust to label noise: A loss correction approach,'' in
  \emph{Proceedings of the IEEE Conference on Computer Vision and Pattern
  Recognition}, 2017, pp. 1944--1952.

\bibitem{ghosh2017robust}
A.~Ghosh, H.~Kumar, and P.~Sastry, ``Robust loss functions under label noise
  for deep neural networks,'' in \emph{Thirty-First AAAI Conference on
  Artificial Intelligence}, 2017.

\bibitem{zhang2018generalized}
Z.~Zhang and M.~Sabuncu, ``Generalized cross entropy loss for training deep
  neural networks with noisy labels,'' in \emph{Advances in neural information
  processing systems}, 2018, pp. 8778--8788.

\bibitem{xu2019l_dmi}
Y.~Xu, P.~Cao, Y.~Kong, and Y.~Wang, ``L\_dmi: An information-theoretic
  noise-robust loss function,'' \emph{arXiv preprint arXiv:1909.03388}, 2019.

\bibitem{reed2014training}
S.~Reed, H.~Lee, D.~Anguelov, C.~Szegedy, D.~Erhan, and A.~Rabinovich,
  ``Training deep neural networks on noisy labels with bootstrapping,''
  \emph{arXiv preprint arXiv:1412.6596}, 2014.

\bibitem{veit2017learning}
A.~Veit, N.~Alldrin, G.~Chechik, I.~Krasin, A.~Gupta, and S.~Belongie,
  ``Learning from noisy large-scale datasets with minimal supervision,'' in
  \emph{Proceedings of the IEEE Conference on Computer Vision and Pattern
  Recognition}, 2017, pp. 839--847.

\bibitem{tanaka2018joint}
D.~Tanaka, D.~Ikami, T.~Yamasaki, and K.~Aizawa, ``Joint optimization framework
  for learning with noisy labels,'' in \emph{Proceedings of the IEEE Conference
  on Computer Vision and Pattern Recognition}, 2018, pp. 5552--5560.

\bibitem{han2019deep}
J.~Han, P.~Luo, and X.~Wang, ``Deep self-learning from noisy labels,'' in
  \emph{Proceedings of the IEEE International Conference on Computer Vision},
  2019, pp. 5138--5147.

\bibitem{han2018co}
B.~Han, Q.~Yao, X.~Yu, G.~Niu, M.~Xu, W.~Hu, I.~Tsang, and M.~Sugiyama,
  ``Co-teaching: Robust training of deep neural networks with extremely noisy
  labels,'' in \emph{Advances in neural information processing systems}, 2018,
  pp. 8527--8537.

\bibitem{chen2019understanding}
P.~Chen, B.~Liao, G.~Chen, and S.~Zhang, ``Understanding and utilizing deep
  neural networks trained with noisy labels,'' \emph{arXiv preprint
  arXiv:1905.05040}, 2019.

\bibitem{sun2019limited}
Y.~Sun, Y.~Tian, Y.~Xu, and J.~Li, ``Limited gradient descent: Learning with
  noisy labels,'' \emph{IEEE Access}, vol.~7, pp. 168\,296--168\,306, 2019.

\bibitem{cheng2019robust}
C.~Xue, Q.~Dou, X.~Shi, H.~Chen, and P.~A. Heng, ``Robust learning at noisy
  labeled medical images: Applied to skin lesion classification,'' in
  \emph{Proceedings - International Symposium on Biomedical Imaging}, 2019, pp.
  1280--1283.

\bibitem{cao2020breast}
Z.~Cao, G.~Yang, Q.~Chen, X.~Chen, and F.~Lv, ``Breast tumor classification
  through learning from noisy labeled ultrasound images,'' in \emph{Medical
  Physics}, 2020, pp. 1048--1057.

\bibitem{xiao2015learning}
T.~Xiao, T.~Xia, Y.~Yang, C.~Huang, and X.~Wang, ``Learning from massive noisy
  labeled data for image classification,'' in \emph{Proceedings of the IEEE
  conference on computer vision and pattern recognition}, 2015, pp. 2691--2699.

\bibitem{shrivastava2016training}
A.~Shrivastava, A.~Gupta, and R.~Girshick, ``Training region-based object
  detectors with online hard example mining,'' in \emph{Proceedings of the IEEE
  conference on computer vision and pattern recognition}, 2016, pp. 761--769.

\bibitem{li2019learning}
J.~Li, Y.~Wong, Q.~Zhao, and M.~S. Kankanhalli, ``Learning to learn from noisy
  labeled data,'' in \emph{Proceedings of the IEEE Conference on Computer
  Vision and Pattern Recognition}, 2019, pp. 5051--5059.

\bibitem{zhang2016understanding}
C.~Zhang, S.~Bengio, M.~Hardt, B.~Recht, and O.~Vinyals, ``Understanding deep
  learning requires rethinking generalization,'' \emph{arXiv preprint
  arXiv:1611.03530}, 2016.

\bibitem{wang2019symmetric}
Y.~Wang, X.~Ma, Z.~Chen, Y.~Luo, J.~Yi, and J.~Bailey, ``Symmetric cross
  entropy for robust learning with noisy labels,'' in \emph{Proceedings of the
  IEEE International Conference on Computer Vision}, 2019, pp. 322--330.

\bibitem{lee2018cleannet}
K.-H. Lee, X.~He, L.~Zhang, and L.~Yang, ``Cleannet: Transfer learning for
  scalable image classifier training with label noise,'' in \emph{Proceedings
  of the IEEE Conference on Computer Vision and Pattern Recognition}, 2018, pp.
  5447--5456.

\bibitem{lee2013pseudo}
D.-H. Lee, ``Pseudo-label: The simple and efficient semi-supervised learning
  method for deep neural networks,'' in \emph{Workshop on Challenges in
  Representation Learning, ICML}, vol.~3, 2013, p.~2.

\bibitem{gong2016teaching}
C.~Gong, D.~Tao, J.~Yang, and W.~Liu, ``Teaching-to-learn and learning-to-teach
  for multi-label propagation,'' in \emph{Thirtieth Aaai Conference on
  Artificial Intelligence}, 2016.

\bibitem{fan2018learning}
Y.~Fan, F.~Tian, T.~Qin, X.-Y. Li, and T.-Y. Liu, ``Learning to teach,''
  \emph{arXiv preprint arXiv:1805.03643}, 2018.

\bibitem{jiang2017mentornet}
L.~Jiang, Z.~Zhou, T.~Leung, L.-J. Li, and L.~Fei-Fei, ``Mentornet: Learning
  data-driven curriculum for very deep neural networks on corrupted labels,''
  \emph{arXiv preprint arXiv:1712.05055}, 2017.

\bibitem{malach2017decoupling}
E.~Malach and S.~Shalev-Shwartz, ``Decoupling" when to update" from" how to
  update",'' in \emph{Advances in Neural Information Processing Systems}, 2017,
  pp. 960--970.

\bibitem{blum1998combining}
A.~Blum and T.~Mitchell, ``Combining labeled and unlabeled data with
  co-training,'' in \emph{Proceedings of the eleventh annual conference on
  Computational learning theory}.\hskip 1em plus 0.5em minus 0.4em\relax ACM,
  1998, pp. 92--100.

\bibitem{he2020momentum}
K.~He, H.~Fan, Y.~Wu, S.~Xie, and R.~Girshick, ``Momentum contrast for
  unsupervised visual representation learning,'' in \emph{Proceedings of the
  IEEE Conference on Computer Vision and Pattern Recognition}, 2020, pp.
  9726--9735.

\bibitem{lin2017focal}
T.~Lin, P.~Goyal, R.~Girshick, K.~He, and P.~Dollár, ``Focal loss for dense
  object detection,'' in \emph{2017 IEEE International Conference on Computer
  Vision}, 2017, pp. 2999--3007.

\bibitem{2009Learning}
A.~Krizhevsky and G.~Hinton, ``Learning multiple layers of features from tiny
  images,'' \emph{Handbook of Systemic Autoimmune Diseases}, vol.~1, no.~4,
  2009.

\bibitem{2017WebVision}
L.~Wen, L.~Wang, L.~Wei, E.~Agustsson, and L.~V. Gool, ``Webvision database:
  Visual learning and understanding from web data,'' 2017.

\bibitem{2017Inception}
C.~Szegedy, S.~Ioffe, V.~Vanhoucke, and A.~A. Alemi, ``Inception-v4,
  inception-resnet and the impact of residual connections on learning,'' 2017.

\bibitem{2020SELF}
D.~T. Nguyen, C.~K. Mummadi, T.~Ngo, T.~Nguyen, L.~Beggel, and T.~Brox, ``Self:
  Learning to filter noisy labels with self-ensembling,'' 2020.

\bibitem{2020Learning}
H.~Song, M.~Kim, D.~Park, and J.~G. Lee, ``Learning from noisy labels with deep
  neural networks: A survey,'' 2020.

\bibitem{2020DivideMix}
J.~Li, R.~Socher, and S.~Hoi, ``Dividemix: Learning with noisy labels as
  semi-supervised learning,'' 2020.

\bibitem{david2019mixmatch}
D.~Berthelot, N.~Carlini, I.~J. Goodfellow, N.~Papernot, A.~Oliver, and
  C.~Raffel, ``Mixmatch: A holistic approach to semi-supervised learning.'' in
  \emph{NeurIPS}, 2019.

\bibitem{toneva2018empirical}
M.~Toneva, A.~Sordoni, R.~T.~d. Combes, A.~Trischler, Y.~Bengio, and G.~J.
  Gordon, ``An empirical study of example forgetting during deep neural network
  learning,'' \emph{arXiv preprint arXiv:1812.05159}, 2018.

\bibitem{laurens2008visiualizing}
V.~Laurens and G.~Hinton, ``Visualizing data using t-sne,'' in \emph{Journal of
  Machine Learning Research}, 2008, pp. 2579--2605.

\end{thebibliography}













\end{document}